\documentclass[]{aa}  
\let\ACMmaketitle=\maketitle
\renewcommand{\maketitle}{\begingroup\let\footnote=\thanks \ACMmaketitle\endgroup}
\usepackage[english]{babel}
\usepackage[utf8]{inputenc}
\usepackage[T1]{fontenc}
\usepackage{graphicx}
\usepackage{amsmath,amssymb}
\usepackage{txfonts}
\usepackage[colorlinks=true,linkcolor=blue,citecolor=blue]{hyperref}
\usepackage{wasysym}
\usepackage{threeparttable,adjustbox}
\usepackage{natbib}
\usepackage{xcolor}
\usepackage{placeins,afterpage}
\usepackage{etoolbox}

\newcommand{\Msun}{$M_{\odot}$}
\newcommand{\Rsun}{$R_{\odot}$}
\newcommand{\Lsun}{$L_{\odot}$}

\newcommand{\kms}{km~s$^{-1}$}

\newcommand{\Ha} {\mbox{H$\alpha$}}
\newcommand{\Hb} {\mbox{H$\beta$}}
\newcommand{\Hg} {\mbox{H$\gamma$}}
\newcommand{\Hd} {\mbox{H$\delta$}}
\newcommand{\He} {\mbox{H$\epsilon$}}
\newcommand{\hei} {\ion{He}{i}}
\newcommand{\Caii} {\ion{Ca}{ii}}
\newcommand{\Fei} {\ion{Fe}{i}}
\newcommand{\Feii} {\ion{Fe}{ii}}
\newcommand{\Nai} {\ion{Na}{i}~D}

\newcommand{\WNT}{WNTR23bzdiq}
\defcitealias{Karambelkar2025ApJ...993..109K}{K25}
\newcommand{\obj}{AT~2025abao}

\begin{document} 

\title{\obj: The fourth luminous red nova in M 31}

\titlerunning{The LRN \obj\ in M 31}

\author{A. Reguitti\inst{\ref{inaf-oapd},\ref{inaf:merate}}\fnmsep\thanks{E-mail: andrea.reguitti@inaf.it},
A. Pastorello\inst{\ref{inaf-oapd}},
G. Valerin\inst{\ref{inaf-oapd}},
F. D. Romanov\inst{3},
A. Siviero\inst{\ref{uni:padua},\ref{inaf-oapd}}, 
Y.-Z. Cai\inst{\ref{china},\ref{china2},\ref{inaf-oapd}},
S. Ciroi\inst{\ref{uni:padua}},
N.~Elias-Rosa\inst{\ref{inaf-oapd},\ref{uni:barcelona}},
T.~Iijima\inst{\ref{inaf-oapd}},
E. Kankare\inst{\ref{uni:turku}}, 
N. Koivisto\inst{\ref{uni:turku}}, 
T. Kravtsov\inst{\ref{uni:turku}}, 
E. Mason\inst{\ref{inaf-oats}},
K. Matilainen\inst{\ref{uni:turku}}, 
A. C. Mura\inst{\ref{uni:padua},\ref{inaf-oapd}}, 
P.~Ochner\inst{\ref{uni:padua},\ref{inaf-oapd}},
T.~M.~Reynolds\inst{\ref{uni:turku},\ref{uni:denmark}},
M.~D.~Stritzinger\inst{\ref{uni:aarhus}}
}

\authorrunning{Reguitti et al.} 

\institute{
INAF – Osservatorio Astronomico di Padova, Vicolo dell'Osservatorio 5, I-35122 Padova, Italy\label{inaf-oapd}
\and
INAF – Osservatorio Astronomico di Brera, Via E. Bianchi 46, I-23807 Merate (LC), Italy\label{inaf:merate}
\and
American Association of Variable Star Observers (AAVSO), 185 Alewife Brook Parkway, Suite 410 Cambridge, MA 02138 USA
\and 
Universit\`a degli Studi di Padova, Dipartimento di Fisica e Astronomia, Vicolo dell'Osservatorio 3, 35122 Padova, Italy\label{uni:padua}
\and
Yunnan Observatories, Chinese Academy of Sciences, Kunming 650216, P.R. China\label{china}
\and
International Centre of Supernovae, Yunnan Key Laboratory, Kunming 650216, P.R. China\label{china2}
\and
Institute of Space Sciences (ICE, CSIC), Campus UAB, Carrer de Can Magrans s/n, 08193 Barcelona, Spain\label{uni:barcelona}
\and 
Department of Physics and Astronomy, University of Turku, 20014 Turku, Finland\label{uni:turku}
\and 
INAF – Osservatorio Astronomico di Trieste, Via G.B. Tiepolo 11, 34143 Trieste, Italy\label{inaf-oats}
\and
Cosmic Dawn Center (DAWN), Niels Bohr Institute, University of Copenhagen, Jagtvej 128, 2200 København N, Denmark\label{uni:denmark}
\and
Department of Physics and Astronomy, Aarhus University, Ny Munkegade 120, DK-8000 Aarhus C, Denmark\label{uni:aarhus}
}

\date{Accepted 24 March 2026. Received 15 February 2026; in original form 7 November 2025}
 
\abstract{
We present photometric and spectroscopic observations of the luminous red nova (LRN) \obj, the fourth discovered in M~31.  
The LRN, associated with the asymptotic giant branch (AGB) star \WNT, was discovered during the fast rise following the minimum phase. It reached its peak at $g=15.1$ mag ($M_g=-9.5\pm0.1$ mag), and then it settled onto a long-duration plateau in the red bands, lasting 70 days, while it was slowly linearly declining in the blue bands.
At the peak the object showed similarities with the canonical LRNe V838 Monocerotis, V1309 Scorpii, and the faint and fast-evolving AT 2019zhd, which is the third LRN in M31, though the later evolution is different.
Spectroscopically, \obj\ evolved as a canonical LRN: the early spectra present a blue continuum with narrow Balmer lines in emission; at the peak, the spectral continuum had cooled to a yellow colour, with a photospheric temperature of 6000 K. Balmer lines had weakened, while absorption lines from metals (\Fei, \Feii, \ion{Sc}{ii}, \ion{Ba}{ii}, \ion{Ti}{ii}) had developed, and they were particularly broad from the UV \Caii\ H\&K lines. Medium- and high-resolution spectra reveal narrow ($\sim$50 \kms) absorption and broad ($\sim$450 \kms) emission profiles in the Balmer lines, from a slower wind and a faster outflow, respectively. Finally, late-time spectra show an orange continuum ($T\sim4000-5000$ K), a return in strength of the Balmer lines and the formation of molecular absorption bands.
\obj\ is the rare case of an LRN with detailed archival information regarding the progenitor system. For the first time, we obtained the spectral energy distribution in the infrared of the precursor of an LRN, which is consistent with that of an M~giant/AGB.
We propose that the dichotomy of light-curve behaviour in LRNe (two peaks vs. plateau) can be explained by the extent and H-richness of the common envelope.
}

\keywords{
Stars: binaries: close – stars: individual: \obj\ – stars: winds, outflows - galaxies: individual: M 31
}

\maketitle

\section{Introduction}\label{introduction}

\begin{figure*}[h]
\includegraphics[width=\textwidth]{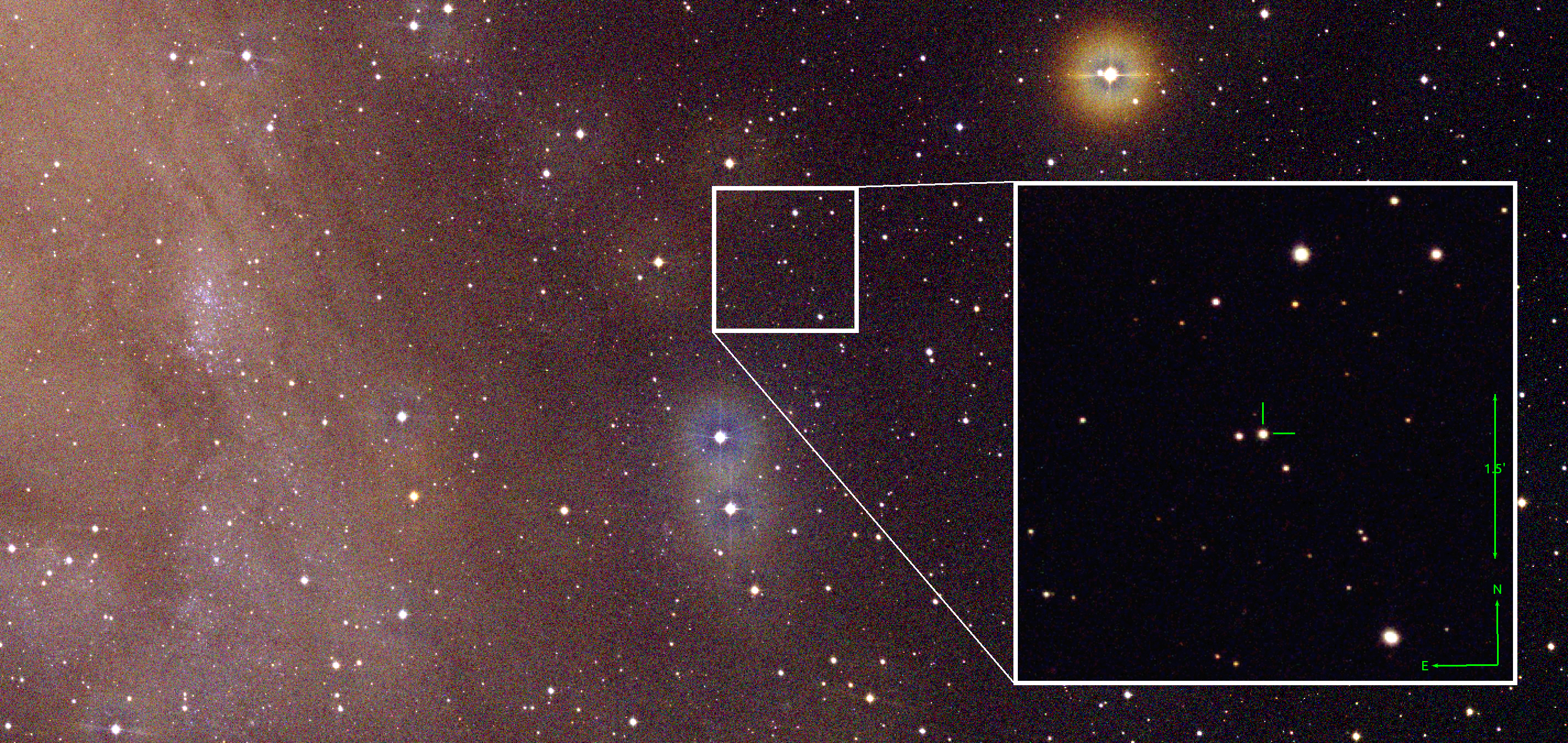}
\caption{Finding chart of \obj. The colour image is a composition of $B$, $V$, and $r$ frames obtained by the Asiago 67/92cm Schmidt telescope on 2025 November 7, at the time of the maximum light. The spiral disc of M 31 is visible towards the left edge of the frame. The transient location is highlighted by the zoomed-in view in the panel to the right. Scale and orientation are reported.
}
\label{fig:chart}
\end{figure*}

Luminous red novae (LRNe; \citealt{Kulkarni2007Natur.447..458K}) are astrophysical transients likely resulting from the coalescence of two non-degenerate stars due to the loss of systemic angular momentum following the ejection of a common envelope \citep[e.g.][]{Tylenda2011A&A...528A.114T, Ivanova2013A&ARv..21...59I, Ivanova2013Sci...339..433I, Kochanek2014MNRAS.443.1319K, Pejcha2014ApJ...788...22P, MacLeod2022ApJ...937...96M}. 
They are a subclass of `interacting gap transients' \citep[IGTs;][]{Pastorello2019NatAs...3..676P, Cai2022Univ....8..493C, Reguitti2026}. IGTs have absolute magnitudes at maximum in the $-15 \lesssim M_V \lesssim -10$ mag range and fill the observational gap that separates the brightest novae from the faintest supernovae. In addition, their photometric and spectroscopic evolution is dominated by the effects of the interaction with a circumstellar medium (CSM; see \citealt{Pastorello2019review} for a review on the phenomenon).

Luminous red novae are identified through a few recurrent observational characteristics. Firstly, they present a light curve with a double peak.
The first peak has a short duration ($\sim$1~week); it is typically very luminous and characterised by a blue colour. This first maximum is followed, a few weeks (or even months) later, by a long-lasting second peak or a plateau-like phase, which is redder and usually fainter than the first.\footnote{However, see the case of AT~2011kp \citep[also known as NGC4490-OT2011;][]{Pastorello2019review, Smith2016}, which had a second peak that was more luminous than the first one.}

The spectra of LRNe show a remarkable evolution, with three main phases: at the time of the first peak, the spectra display a blue continuum with narrow Balmer emission lines, similar to those of SNe IIn \citep{Fraser2020RSOS....700467F}.
At the second peak, which sometimes is more akin to a plateau, the continuum cools to that of a yellow or orange star ($\sim$5000--6000 K). At this phase, the Balmer lines become very weak, while a forest of absorption lines from neutral and singly ionised metals (such as O, Ca, Fe, Sc, Ba, V, and Ti) dominate the spectrum.
Finally, in the third stage -- after the second peak -- the spectral continuum changes again to resemble that of a red giant star, with a photospheric temperature of only 3000--4000 K. On top of it, the narrow \Ha line in emission becomes prominent again, whereas molecular absorption bands from VO and TiO devour the continuum.

For a number of extragalactic LRNe, the progenitor systems were detected in pre-outburst archival images from deep ground-based facilities or space telescopes \citep{Fraser2011ATel.3574....1F, Mauerhan2015MNRAS.447.1922M, Smith2016, Blagorodnova2017, Cai2019A&A...632L...6C, Pastorello2021_20hat, Blagorodnova2021, Cai2022A&A...667A...4C}.
From these findings, it was discovered that the typical progenitors of LRNe are yellow hypergiant stars, with masses up to 50~\Msun\,\citep{Guidolin2025}, and they are suggested to sit in the Hertzsprung gap of the HR diagram \citep{Tranin2025A&A...695A.226T, Addison2022MNRAS.517.1884A}.

Most LRNe are discovered in distant galaxies; hence, they are usually faint targets, making their observational campaigns challenging and, consequently, limiting the amount of information we can infer from their study. For this reason, it is important to find and study IGTs in nearby galaxies, especially within the Local Group.
A handful of LRNe were observed in the Milky Way (MW): V4332~Sgr \citep{Martini1999AJ....118.1034M, Tylenda2005_V4332}, OGLE-2002-BLG360 \citep{Tylenda2013A&A...555A..16T}, V838~Mon \citep{Munari2002A&A...389L..51M, Bond2003Natur.422..405B, Tylenda2005_V838}, and V1309~Sco \citep{Mason2010A&A...516A.108M, Tylenda2011A&A...528A.114T}. 
Three others were instead observed in M~31: M31-RV \citep{Mould1990ApJ...353L..35M, Boschi2004AA...418..869B}, M31-LRN2015, \citep{Kurtenkov2015A&A...578L..10K, Williams2015ApJ...805L..18W, MacLeod2017ApJ...835..282M, Lipunov2017MNRAS.470.2339L, Blagorodnova2020}, and AT~2019zhd \citep{Pastorello2021_19zhd}.

In this work, we present the results of our follow-up campaign of \obj, the fourth event of its type discovered in the Andromeda galaxy.
The proximity of M 31 (only 750~kpc away) allowed us to constrain the progenitor parameters, and to provide a rich dataset of the transient event in terms of cadence, time coverage, resolution, and signal. Another paper was published illustrating the pre-outburst evolution of the source that later generated the LRN \obj\ \citep[labelled with the survey names WNTR23bzdiq and WTP19aalzlk;][hereafter \citetalias{Karambelkar2025ApJ...993..109K}]{Karambelkar2025ApJ...993..109K}, and predicting a later evolution towards an LRN. In particular, the discovery of \obj\ offered the rare opportunity to perform a detailed analysis of its progenitor system in a stage of relative quiescence and to study its spectro-photometric properties a short time before the outburst onset in detail.\footnote{This phase was well monitored in photometry only for the Galactic LRN V1309~Sco \citep{Tylenda2011A&A...528A.114T}.}
Through the analysis of infrared data, \citetalias{Karambelkar2025ApJ...993..109K} inferred that the progenitor of \obj\ was an early asymptotic giant branch (AGB) star with a temperature of about 3,500~K, a luminosity of $1.6 \times 10^4$ \Lsun, a radius of 350 \Rsun, and a mass of $7\pm2$~\Msun. Over a period of about seven years before the LRN outburst, the star increased its luminosity by a factor of three, with a limited colour and/or temperature evolution, while the spectrum showed properties compatible with those of an M-type star. The global (though slow) luminosity rise of the source was accompanied by fluctuations in its light curve, compatible with the presence of a binary companion. 
\citetalias{Karambelkar2025ApJ...993..109K} suggested that the stellar brightening was the consequence of the common envelope evolution of a binary system having an AGB star as a primary member. This paper analyses the later evolution of the object, from the late pre-LRN phases to the LRN outburst.

The structure of the paper is as follows: in Section \ref{discovery} we present the discovery circumstances of \obj\ and some observational characteristics. The photometric dataset and the light-curve evolution are presented in Section \ref{photometry}, where the transient is also compared to other LRNe objects with similar properties. The spectral sequence is shown and analysed in Section~\ref{spectroscopy}, and finally the results are discussed in Section \ref{discussion}.

\section{Discovery}\label{discovery}

The LRN \obj\ was officially discovered on 2025 October 19\footnote{Universal time is used throughout the paper.}, or modified Julian date (MJD) = 60967.8196, by the Mobile Astronomical System of Telescope-Robots program \citep{Kechin2025TNSTR4205....1K}, at an unfiltered magnitude of 17.3 (Vega mag), and immediately reported to the Transient Name Server\footnote{https://www.wis-tns.org/} (TNS). However, the first sighting of a brightening from a known transient was made by Koichi Itagaki on October 17.4798, at an unfiltered magnitude of 17.8 (Vegamag).\footnote{http://www.cbat.eps.harvard.edu/unconf/followups/J00384865+4046079.html} An even earlier pre-discovery measurement was provided by \cite{Fabregat2025ATel17527....1F} on October 14, at $V=$18.6 mag.
The LRN appeared in the south-west quadrant of the very nearby Andromeda galaxy (M~31), at celestial coordinates $\alpha =$ 00:38:48.62, $\delta = +$40:46:07.6 ($45'$ west and $30'$ south of the nucleus), in the outskirts of its spiral disc. The location of the transient is shown in Fig.~\ref{fig:chart}.
\obj\ was correctly classified on 2025 November 2 as an LRN by \cite{Taguchi2025TNSCR4414....1T}, who noted similarities with another LRN found in M~31, AT~2019zhd \citep{Pastorello2021_19zhd}. 
For the distance to M~31, we assumed the same value adopted by \cite{Pastorello2021_19zhd} ($\mu=24.47\pm0.06$ mag, $d=0.785$ Mpc). Its redshift is $z = - 0.001$ \citep{Falco1999PASP..111..438F}. However, we note that the peaks of the spectral lines of \obj\ are blueshifted by $z = - 0.0015$, which we adopt hereafter. The additional radial velocity is likely caused by the rotation curve of the galaxy disc, with the object being located on the side approaching us at $\sim$150 \kms\ \citep{Chemin2009ApJ...705.1395C}.
The Galactic reddening towards \obj\ is modest, although not negligible: $A_V=0.17$ mag \citep{Schlafly2011ApJ...737..103S}. Given the peripheral location of \obj\ in the disc of M~31, any additional contribution to the dust extinction is likely negligible. Hereafter, in agreement with \citetalias{Karambelkar2025ApJ...993..109K}, we only account for the dust attenuation within the MW.

\section{Photometric evolution}\label{photometry}
\subsection{Observations and data reduction}

\begin{figure*}\centering
\includegraphics[width=1.95\columnwidth]{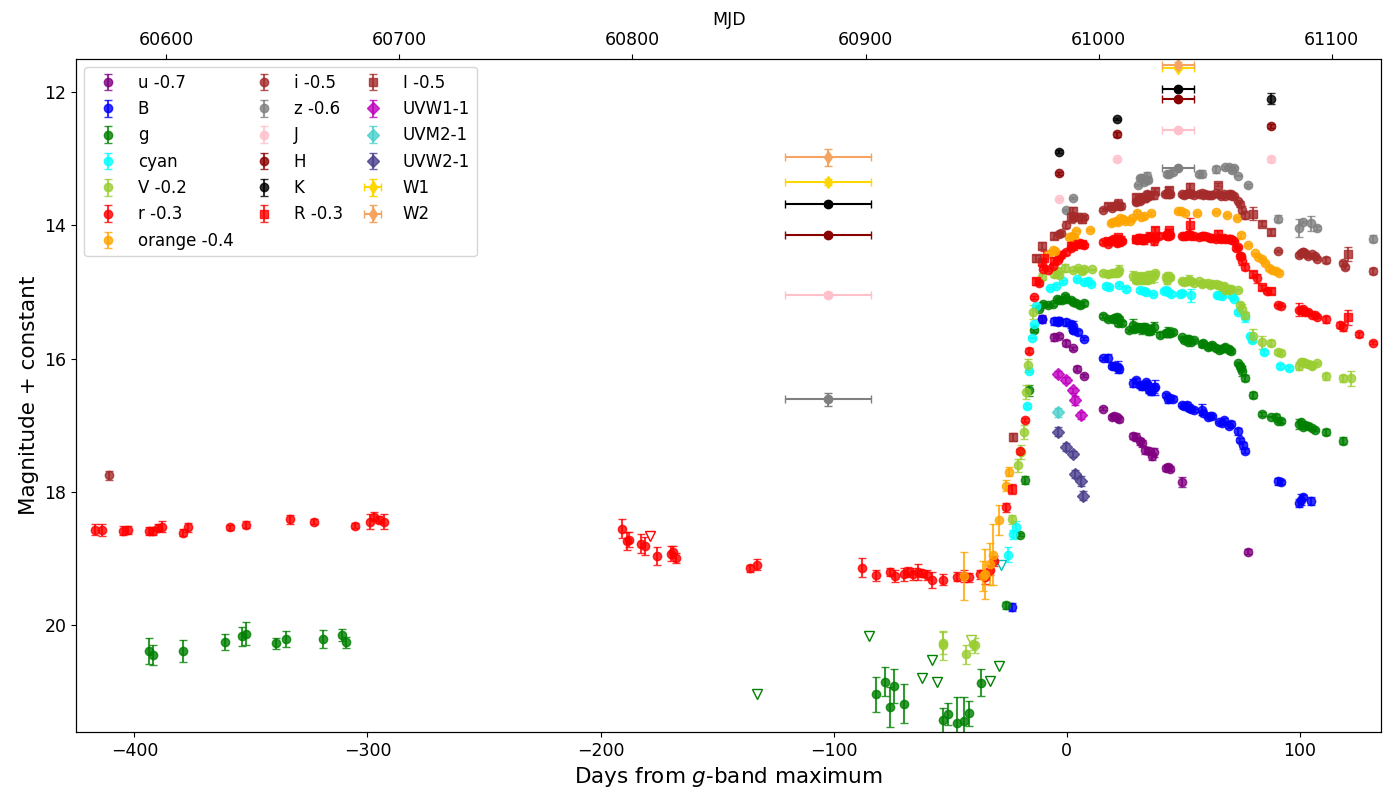}
\includegraphics[width=1.95\columnwidth]{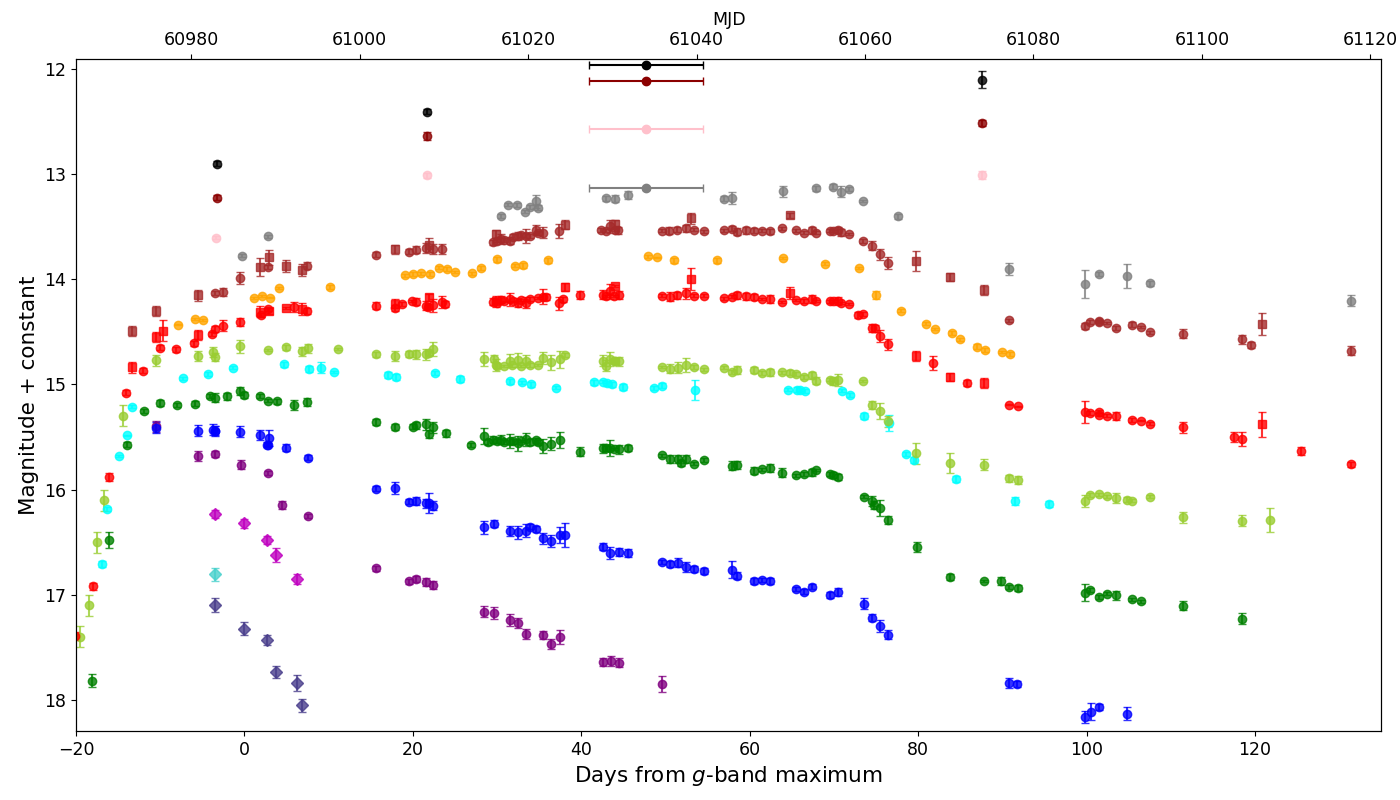}
\caption{Optical light curves of \obj. Top panel: the entire evolution since October 2024, when the follow-up campaign of \WNT\ conducted by \citetalias{Karambelkar2025ApJ...993..109K} ended. The pre-LRN light curve, its decline to the minimum, and the final rapid rise to the LRN peak are visible.
The temporal uncertainty in the spectrophotometry data from \textit{SPHEREx} is due to the datapoints being collected across a time interval of about 5 and 2 weeks. Hence, the magnitudes are shown as taken at the mid-time of the interval, and with an error bar on the time.
Bottom panel: zoom in on the fast rise towards the maximum light, the 70 days-long plateau, the fall from it, and the following linear decline.}
\label{fig:light_curve}
\end{figure*}

While \citetalias{Karambelkar2025ApJ...993..109K} presented early photometric data of \obj, this work complements their light curve with previously unpublished pre-LRN photometric points from the Zwicky Transient Facility \citep[ZTF;][]{Bellm2019PASP..131a8002B, Graham2019PASP..131g8001G}.
The initial phases following the LRN outburst were exclusively monitored by amateur astronomers and all-sky surveys, such as ZTF and the Asteroid Terrestrial-impact Last Alert System (ATLAS; \citealt{Tonry2018PASP..130f4505T}).
Our multi-band optical and near-infrared (NIR) follow-up campaign began shortly after the spectroscopic classification of the object and lasted for approximately four months. The \textit{Swift} space telescope was also triggered to obtain ultraviolet (UV) photometry at early phases. The facilities used to collect our photometric data are listed in Table \ref{tab1}.
The photometric data were reduced using standard procedures with the dedicated \texttt{ECsnoopy} pipeline.\footnote{\texttt{ECsnoopy} is a package for SN photometry using PSF fitting and/or template subtraction developed by E. Cappellaro. A package description can be found at \url{https://sngroup.oapd.inaf.it/ecsnoopy.html}.}
Instrumental magnitudes were determined through the point spread function (PSF) fitting technique, and calibrated against the catalogue of the Sloan Digital Sky Survey (SDSS) \citep{Ahumada2020ApJS..249....3A}.
The \textit{Swift/UVOT} UV data were reduced with the HEASOFT pipeline v.~6.35.2\footnote{NASA High Energy Astrophysics Science Archive Research Center – Heasarc 2014.} by aperture photometry within a fixed radius of 5".
We also retrieved calibrated forced photometry measurements from the ALeRCE Broker \citep{Forster2021AJ....161..242F} for ZTF and from the public Forced Photometry server\footnote{\url{https://fallingstar-data.com/forcedphot/}} \citep{Shingles2021TNSAN...7....1S} for ATLAS.
Given its apparent brightness, the object was also observed extensively by amateur astronomers;\footnote{More observations of \obj\ during the outburst are presented by \cite{Mikolajczyk2026TNSAN..21....1M}.} the magnitudes were measured via aperture photometry on their pre-reduced frames (corrected for bias, flat-field, and dark frames), and calibrated their measurements against the AAVSO Photometric All-Sky Survey\footnote{\url{https://www.aavso.org/apass}} catalogue \citep{APASS2019JAVSO..47..130H}.
The final UV, optical (Sloan, Johnson, and ATLAS), and NIR magnitudes are listed in the CDS, while the light curves are plotted in Fig.~\ref{fig:light_curve}.

\subsection{Light curve and its interpretation}

\subsubsection{Pre-outburst phase} \label{sect:photopreoutburst}
The photometric evolution of the transient during the pre-LRN was extensively studied by \citetalias{Karambelkar2025ApJ...993..109K}. The light curve of the transient experienced a slow, long-lasting (about seven years) brightening, with some luminosity fluctuations superposed on that global trend. Additional photometric points presented in this paper show that the source reached a relative maximum about one year before the LRN discovery, followed by a poorly-sampled luminosity decline to a pre-outburst minimum (see also Sect.~\ref{Sect:comparisons}).

The pre-outburst evolution of \obj, from the slow luminosity rise to a broad peak, superposed magnitude fluctuations, and ensuing luminosity decline, closely resembles that observed in V1309~Sco \citep{Tylenda2011A&A...528A.114T}. In both cases, the peculiar light-curve evolution is likely a consequence of unstable mass transfer between the stellar components of a binary system \citep[Roche-lobe overflow; e.g.][]{Pejcha2014ApJ...788...22P, Pejcha2016MNRAS.455.4351P}. This culminates with the formation of a common envelope embedding the stellar system, which becomes progressively optically thick \citep{Ivanova2013A&ARv..21...59I}.
A slow pre-LRN luminosity rise was also observed in other LRNe discovered in M~31, such as M31-LRN-2015 \citep{Blagorodnova2020} and AT~2019zhd \citep{Pastorello2021_19zhd}, as well as in several extragalactic LRNe, although with some limitations due to a lower monitoring cadence and the poorer signal-to-noise datasets.\footnote{The slow brightenings heralding LRN eruptions were observed for AT~2014ib \citep[SNhunt248;][]{Kankare2015A&A...581L...4K, Mauerhan2015MNRAS.447.1922M}, AT~2015dl \citep[M101-2015OT1;][]{Blagorodnova2017,Goranskij2016AstBu..71...82G}, AT~2020hat \citep{Pastorello2021_20hat, Reguitti2026}, AT~2021blu \citep{Pastorello2023} and, more marginally, in AT~2011kp \citep{Pastorello2019review} and AT~2021biy \citep{Cai2022A&A...667A...4C}.}

\subsubsection{LRN outburst}

In the case of \obj, this phase is followed by a fast brightening of about five magnitudes in the $r$ band (6 in $g$); this is the onset of the LRN outburst. The rapid rise lasts about one week from the discovery of \obj. Then, its light curve settles on a sort of plateau phase during which the object maintains a nearly constant brightness ($V\simeq 14.9$ mag). To constrain the timing of the main LRN maximum, we performed a second-order polynomial fit to the $g$-band data collected between MJD = 60973 and MJD = 61000, determining that the light curve peaked on MJD 60986.3$\pm$0.2, at a magnitude of $g=15.10\pm0.02$ mag. 
We used this epoch as a reference for all phases considered in this paper. We note that the bluer band light curves peak earlier ($-2.7$~d in $u$, $-1.4$~d in $B$), while a photometric peak is not well-defined in the red bands, where a plateau is noticeable, lasting for about two months.
We measured the post-maximum decline rates in the optical bands by fitting the light curve in each filter with a straight line. The decline is steepest in the $u$ band ($3.9 \pm 0.2$ mag 100 d$^{-1}$), and less and less steep as we move towards redder filters ($\gamma_B=2.8 \pm 0.1$, $\gamma_g=1.5 \pm 0.1$, $\gamma_V=0.29 \pm 0.03$ mag 100 d$^{-1}$). The $r$-band light curve instead is flat, and the $i$-band one is actually still rising, up to phase $+45$~d, when it reached its maximum.
The plateau phase ends at $+72$~d; it is followed by a sudden fall of the light curve and, 20 days later, by a linear decline with a slope of $\gamma_r=1.18 \pm 0.06$ mag~100~d$^{-1}$.
 
Individual patterns of LRN light curves are not fully understood. However, there is a general consensus that the characteristic double-peaked shape is due to a merging event \citep{Banerjee2026ApJ...999L..35B}.
As a consequence of the common envelope ejection and the loss of systemic angular momentum, the two components of the binary system may merge. If this happens, the main LRN outburst is produced by the violent ejection of material following the coalescence of the two stellar cores, and its subsequent interaction with the pre-existing common envelope \citep[e.g.][]{Metzger2017MNRAS.471.3200M, MacLeod2017ApJ...835..282M, MacLeod2022ApJ...937...96M, Matsumoto2022ApJ...938....5M, Kirilov2025ApJ...994L..41K, Chen2026ApJ...999..217C}.

\subsection{Colour evolution}
\begin{figure}
\includegraphics[width=1\columnwidth]{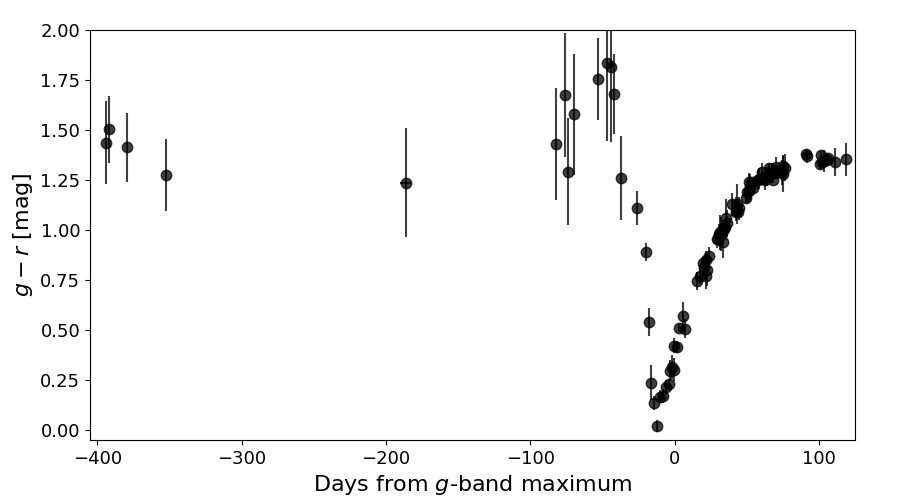}
\includegraphics[width=1\columnwidth]{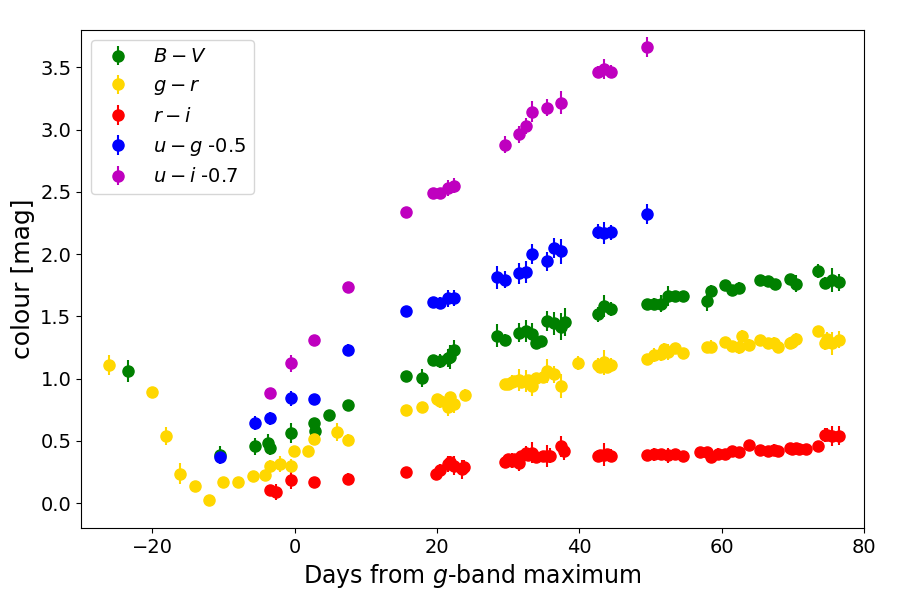}
\caption{Colour curves of \obj. Top panel: Evolution of the $g-r$ colour from $-1.1$ yr to $+3$ months. The colour changes prominently between the slow pre-outburst rise and optically thick phase and the fast rise and the first peak. 
Bottom panel: Evolution of multiple colour curves ($B-V$, $g-r$, $r-i$, $u-g$, $u-i$) around the maximum light and plateau.
}
\label{fig:colors}
\end{figure}

The intrinsic colours of \obj, which are blue soon after its discovery ($B-V=0.45$ mag at $-4$~d), evolve to redder colours after the $V$-band maximum ($B-V=0.8$ mag at $+8$~d, $B-V=1.0$ mag at $+15$~d).
Thanks to the scanning by the ZTF survey we have, for the first time, an indication of the optical colour of an LRN during the pre-outburst slow luminosity rise. The $g-r$ colour stays roughly around 1.4 mag one year before the $g$-band maximum. Then, it reaches a maximum up to $\approx1.7$ mag (with large scatter) at the time of the $r$-band minimum ($-2$ months). $g-r$ decreases to 1.1 mag at $-30$ d, at the start of the fast rise to the first peak, diminishing to a lowest of $+0.0$ mag at $-12$ d. From there, $g-r$ increases again to 1.3 mag at $+76$~d, at the time of the plateau drop. 
The colour curves of \obj\ are shown in Fig.~\ref{fig:colors}.

\subsection{Comparisons with similar objects}\label{Sect:comparisons}
\begin{figure}
\includegraphics[width=1\columnwidth]{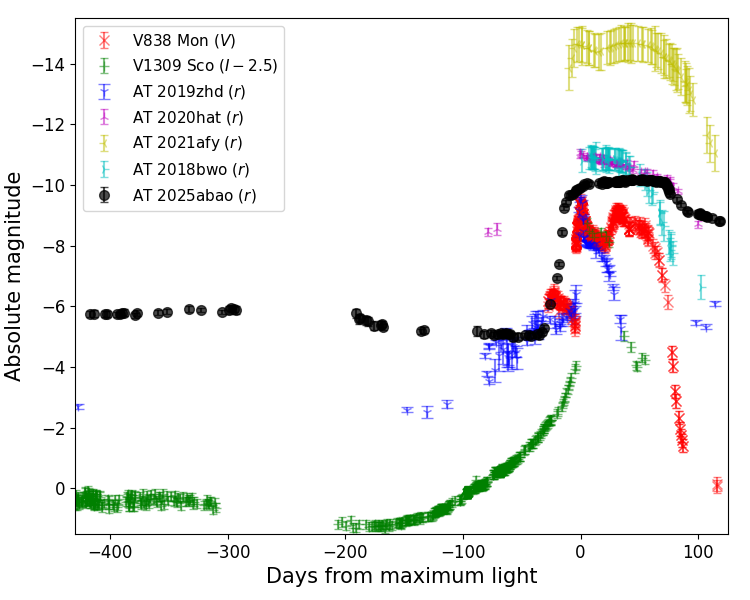}
\caption{Absolute light curve of \obj in the $r$ band, compared with those of AT~2019zhd, V838~Mon ($V$-band), V1309~Sco ($I$-band), AT 2018bwo, AT 2020hat, and AT 2021afy. The light curve of V1309~Sco is scaled upwards by 2.5 magnitudes to match the peak brightness and to see the pre-outburst slow rise of both V1309~Sco and \obj. The error bar on the distance modulus of V1309 Sco (0.724 mag) is not shown.
}
\label{fig:absolute}
\end{figure}

As comparison objects, we select a few faint LRNe well monitored during the slow pre-outburst rise phase. Given that \obj\ has an extensive and well-sampled light curve since a few years before the LRN event \citepalias{Karambelkar2025ApJ...993..109K}, our primary reference object is V1309~Sco \citep{Mason2010A&A...516A.108M, Tylenda2011A&A...528A.114T}, although the two objects have significantly different light curves during the outburst.
\obj\ reaches an absolute magnitude at the $r$-band peak of $M_r=-10.2\pm0.1$ mag (Fig.~\ref{fig:absolute}), hence it lies at the faint edge of the extragalactic LRNe brightness distribution, although it is significantly more luminous than Galactic objects such as V1309~Sco and V4332~Sgr \citep{Pastorello2019review}.
AT~2019zhd \citep[in M~31;][]{Pastorello2021_19zhd} and the Galactic V838~Mon (\citealt{Goranskij2020AstBu..75..325G} and references therein) have similar absolute magnitudes to \obj\ at their maximum, and they were hence selected as comparison objects.

As mentioned in Sect. \ref{sect:photopreoutburst}, the pre-outburst light curves of both \obj\ and V1309~Sco were very well monitored  (see Fig. \ref{fig:absolute}). While the former was much brighter at the top of the slowly rising phase (about one year before the LRN outburst onset), reaching $M_r\sim -6$ mag, the latter reached only $M_I\sim +3$ mag. Then, both objects dimmed towards the optically thick common-envelope phase. In particular, \obj\ reached a minimum at $M_r=-5.0$ mag at $-42$~d, before starting the fast rise towards the LRN maximum light, when it increased its brightness by about 6 mag.

From the comparison in Fig. \ref{fig:absolute}, we note that the fast rise of \obj\ to the main peak is very similar to that exhibited by AT~2019zhd in terms of both speed and luminosity. At the peak, the absolute magnitude of \obj\ in the $r$ band is only marginally brighter than those of AT~2019zhd ($M_r=-9.6$ mag) and V838~Mon ($M_R=-9.9$ mag; note that the figure shows the $V$-band light curve, which peaks at $M_V=-9.6$ mag).
Then, its light curve shows a flat, long-lasting (about two months) plateau, while the comparison objects show a short-duration (less than a week) early peak before declining. V838~Mon shows a more structured post-maximum light curve, with a later re-brightening \citep[e.g.][]{Munari2002A&A...389L..51M}.
In contrast to the comparison objects, the outburst of V1309~Sco is much fainter, and a shift upwards by 2.5 mag has to be applied to its $I$-band absolute light curve in Fig. \ref{fig:absolute} to match the peak luminosity of \obj.

Due to the plateau-like feature after the first peak, the light curve of \obj\ is reminiscent of other LRNe that do not exhibit an evident second red peak, instead showing a long phase of nearly constant luminosity. Known LRNe with plateau-like light curves are AT~2021afy \citep{Pastorello2023}, although it is 4.5 mag more luminous and at the bright edge of the LRNe brightness distribution; AT~2018bwo, whose plateau reaches $M_r=-10.8$ mag \citep{Blagorodnova2021, Pastorello2023}; and AT~2020hat, which peaks at $M_r=-11$ mag \citep{Pastorello2021_20hat}. However, we remark that AT~2018bwo was already discovered after the first peak due to the solar conjunction; hence, the object could have reached a higher peak luminosity.

\subsection{Bolometric light curve}
\begin{figure}
\includegraphics[width=1\columnwidth]{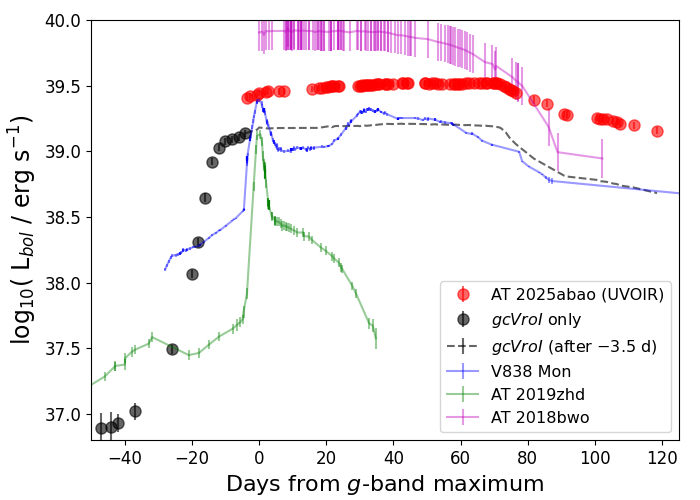}
\caption{Pseudo-bolometric (just the $g, c, V, r, o, I$ bands) light curve of \obj, in black; and UV+Optical+NIR ($UVW2$ to $K$ bands) bolometric curve of the main outburst, in red.
The separation is made at phase $-3.5$ d, when our multi-band photometric follow-up campaign started (after this phase, the $gcVroI$ pseudo-bolometric is shown as a dashed line).
While the bolometric luminosity of \obj\ was similar to those of the comparison objects at the time of the $g$-band maximum, the subsequent evolution is different.
}
\label{fig:bolom}
\end{figure}

Given that only some photometric bands are available before the discovery and the start of the follow-up campaign of the main outburst, we divided the dataset into two parts and constructed two different bolometric light curves. 
We first constructed a pseudo-bolometric curve selecting only the $g, c, V, r, o,$ and $I$ bands, integrating their fluxes (inferred by converting the magnitudes taking into account for the distance and reddening) with the trapezoidal rule from $-50$ to $-3.5$ days, hence covering the pre-discovery phases. Then, we built the bolometric curve of \obj\ extended from the UV to the NIR, accounting for the contribution from $UVW2$ to the $K$ bands, starting at the discovery epoch, hence only covering the main LRN outburst. 
For epochs without observations in some bands, we interpolated the available data using the $r$-band light curve as a reference and assuming a constant colour index.
Both the pseudo- and the fully bolometric curves of \obj\ are shown in Fig. \ref{fig:bolom}. 

The bolometric luminosity of \obj\ at the time of the $g$-band maximum is $\rm log_{10}(L_{bol}/\rm{erg\, s^{-1}})=39.44\pm0.03$, similar to the values reached by the comparison objects. However, its top luminosity ($\rm log_{10}(L_{bol}/\rm{erg\, s^{-1}})=39.52\pm0.03$) was reached about 67~d later, just before the fall from the plateau.
This peak bolometric luminosity makes \obj\ the brightest LRN ever discovered in M~31 and, at the time of the peak, one of the most luminous objects within that galaxy.
However, while the bolometric luminosity of \obj\ at the peak is similar to that of the comparison objects and the fast rise resembles that of AT~2019zhd, the later evolution is completely different, as none of the LRNe considered present a plateau lasting 70 days. In this respect, \obj\ is more similar to AT~2018bwo \citep{Blagorodnova2021, Pastorello2023}, although the former is about three times fainter.
The total radiated energy, calculated by integrating the UV+optical+NIR bolometric curve between phases $-3.5$ and $+118$ d, is $(2.86\pm0.07_{stat}\pm0.03_{syst})\times10^{46}$ erg. This value has to be considered as a lower limit since we are missing the contribution emitted at wavelengths longer than 2.5 $\mu$m and shorter than 2000 \AA, and prior to phase $-3.5$~d, i.e. at the time of the fast rise before the maximum light.
If we add the contribution of $u$ to $I$ bands between phases $-15$ and $-3.5$ d, and of the $gcVroI$ filters between $-50$ and $-15$ d, we obtain a total radiated energy of $\sim3.1\times10^{46}$ erg.

\subsection{SED evolution} \label{Sect:SED}
\begin{figure}
\includegraphics[width=1\columnwidth]{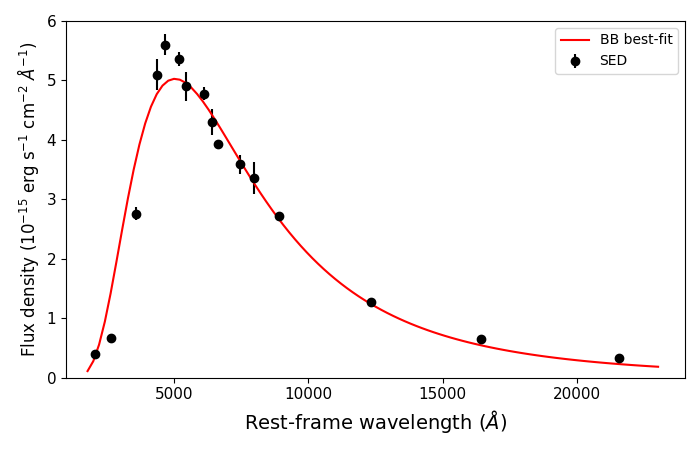}
\caption{SED of \obj\ at the epoch of the $g$-band maximum, together with the BB best-fit. A single Planckian function is sufficient to obtain a good match. Its parameters are $T_{\rm BB}=5740\pm 140$ K and $R_{\rm BB}=870\pm 50$ \Rsun.
}
\label{fig:SED}
\end{figure}

\begin{figure}
\includegraphics[width=1\columnwidth]{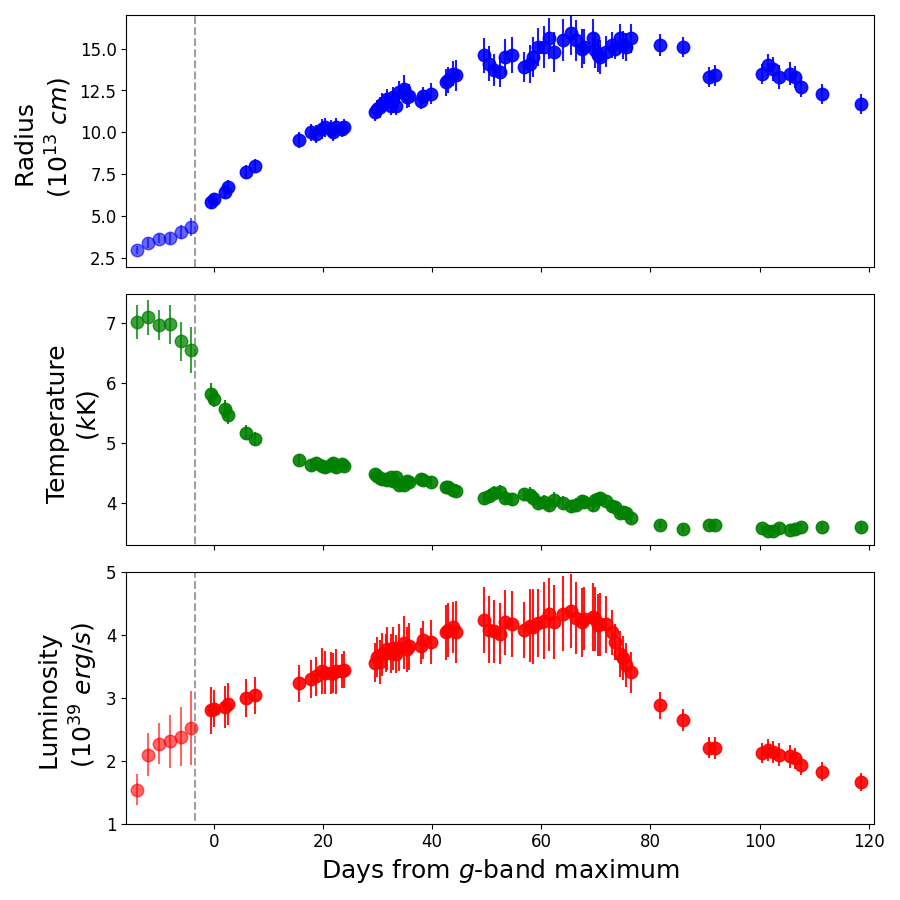}
\caption{Evolution of the BB radius, temperature, and luminosity of \obj\ starting from phase $-15$ d. Before $-3.5$ d (epoch marked by the dashed grey line), the parameters are determined based on only the filters between $u$ and $I$, and afterwards they are estimated from the UV to the NIR coverage.
}
\label{fig:RT}
\end{figure}

We constructed the spectral energy distribution (SED) of \obj\ at different epochs, starting from phase $-3.5$~d, combining the photometric data from the NUV to the NIR domains. Then, we fitted them with a Planckian function to estimate how the black-body (BB) temperature ($T_{\rm BB}$) and the radius ($R_{\rm BB}$) evolve with time. 
Between $-15$ d and $-3.5$ d, we did so using only the data from the filters from $u$ to $I$. For epochs without the photometric information in a given filter, we made interpolations or extrapolations from adjacent epochs with available data, assuming no colour changes.
The highest temperature was reached on $-12$ d (the epoch of the minimum $g-r$ colour), at $T_{\rm BB}=7100\pm300$ K. At the time of the $g$-band maximum light (see Fig.~\ref{fig:SED}), the SED peaks at around 5000~\AA\ and is fitted with a single BB function that has a temperature of $5740\pm140$~K. The inferred photospheric radius is $870\pm50$ \Rsun. At these early phases, no IR excess is detected. In Fig.~\ref{fig:RT}, we plot the temporal evolution of $T_{\rm BB}$, $R_{\rm BB}$, and the BB luminosity $L_{\rm BB}$.
During the plateau phase, the temperature slowly decreases and flattens at $T_{\rm BB}\approx$4300~K, slightly lower than expected for the H recombination, as also evidenced by the redder shape of the spectral continuum in this phase (see Sect. \ref{spectroscopy}).
The photospheric radius when the object was hottest is $(3.4\pm0.3)\times10^{13}$ cm (corresponding to $500\pm40$~\Rsun), while during the plateau phase it slowly expands to $\sim(15\pm1)\times10^{13}$ cm (hence, $2150\pm140$ \Rsun).
During the plateau phase, $L_{\rm BB}$ slightly increases from $3\times10^{39}$ at the time of the $g$-band maximum to $4\times10^{39}$ erg~s$^{-1}$ about 70~d later, before a sharp decrease occurs after the fall from the plateau.
    
\subsection{\textit{SPHEREx} observations}
\subsubsection{SED of the precursor}
\begin{figure}
\includegraphics[width=1\columnwidth]{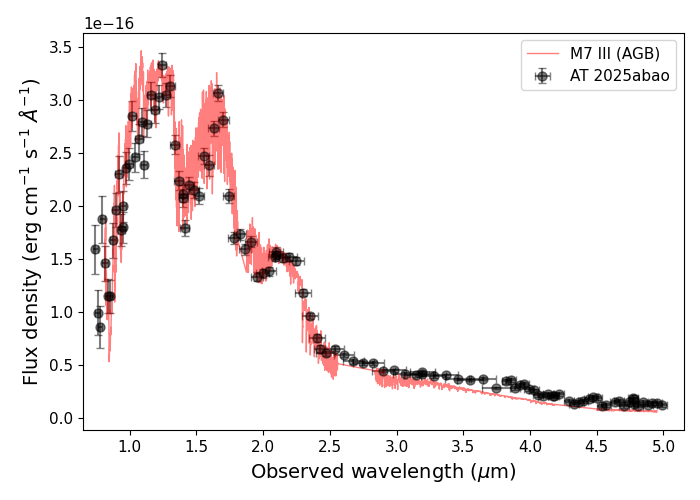}
\caption{Infrared SED of the precursor of \obj\ between 0.74 and 5.0 $\mu$m as obtained by SPHEREx 3--4 months before the $g$-band maximum light. 
Overplotted is a spectrum of an M7e III/AGB star.
} 
\label{fig:Spherex}
\end{figure}

The Spectro-Photometer for the History of the Universe, Epoch of Reionization and Ices Explorer \citep[SPHEREx;][]{Bock2026ApJ...999..139B} is a space telescope launched on 2025 March 11 to conduct an all-sky survey in 102 infrared colours between 0.75 and 5 $\mu$m, hence covering a wide range of wavelengths from the NIR to the mid-infrared (MIR) domains.
During its first scanning of the whole celestial sphere, between MJD 60865 and 60902 ($-121$~d to $-84$ d), it observed the field of \obj\ during the pre-LRN phase, at a time close to the optical minimum.
We retrieved the public calibrated SED \citep{Akeson2025arXiv251115823A, Hui2026arXiv260209139H} from the spectrophotometry tool available on the NASA/IPAC Infrared Science Archive web page.\footnote{\url{https://irsa.ipac.caltech.edu/applications/spherex/tool-spectrophotometry}}
The IR SED of the precursor of \obj\ from SPHEREx is presented in Fig.~\ref{fig:Spherex}.

This provides a rare opportunity to analyse an NIR+MIR spectrum of a precursor of an LRN, and the MIR region is observed in spectroscopy for the first time.
The source is faint in the bluest bands and shows a red continuum between 0.74 and 1.2 $\mu$m.
Three peaks are clearly present at 1.3, 1.7, and 2.2~$\mu$m. These peaks, and the broad absorptions in between due to molecular absorption bands from TiO, H$_2$O, and CO, are also visible in the NIR spectra presented by \citetalias{Karambelkar2025ApJ...993..109K} and taken in 2024 (see their Figure 3).
Then, at $\lambda>2.5\ \mu$m, the SED is nearly flat in the MIR, which could be due to thermal emission from dust surrounding the progenitor system of \obj. 
For reference, in Fig.~\ref{fig:Spherex} we plot an NIR+MIR spectrum of a M7e III star (HD 108849, a known AGB\footnote{\url{https://simbad.u-strasbg.fr/simbad/sim-basic?Ident=bk+vir}}) from the Atlas by \cite{Rayner2009ApJS..185..289R}, which perfectly matches the SED from SPHEREx. A modelling of the SED is presented in Sect. \ref{discussion}.
Finally, we performed spectro-photometry on the SED to derive approximate NIR and MIR magnitudes (in $z$, $J$, $H$, $K$, WISE $W1$ and $W2$ bands) of \obj\ 3--4 months before the $g$-band maximum. The infrared colours are larger (thus redder) than during the outburst (i.e. $J-K=1.36\pm0.05$ mag vs. $0.61\pm0.03$ mag at $+22$ d).

\subsubsection{SED of the outburst}
\begin{figure}
\includegraphics[width=1\columnwidth]{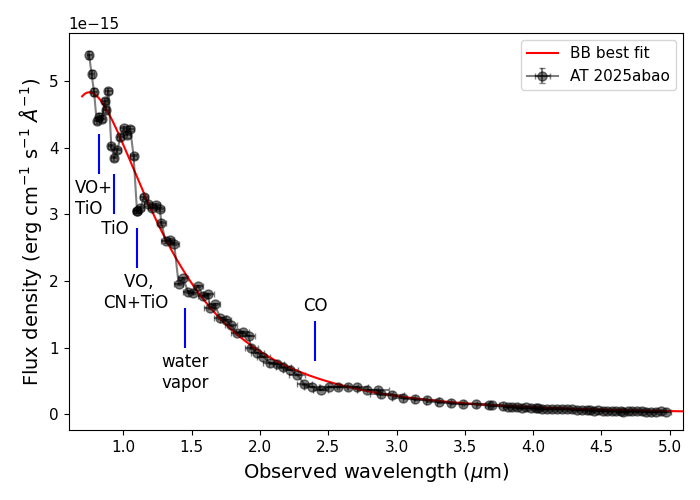}
\caption{Infrared SED of \obj\ from SPHEREx at around $+$1.5 months. The BB best-fit is also shown, while the main molecular absorption features are marked.
} 
\label{fig:Spherex_post}
\end{figure}

Between MJD 61027 and 61041 ($+41$ d to $+55$ d), SPHEREx observed the field of \obj\ again, this time during the outburst. The corresponding IR SED, presented in Fig.~\ref{fig:Spherex_post}, shows a hot thermal continuum, with absorption bands at 0.82 $\mu$m (analogous to those seen in the optical spectra, produced by VO and TiO molecules, as also observed in V838 Mon; \citealt{Lynch2004ApJ...607..460L}), 0.94 $\mu$m \citep[due to TiO;][]{Valenti1998ApJ...498..851V}, 1.1 $\mu$m \citep[from a combination of VO, CN, and TiO molecules;][]{Pastorello2023}, 1.45 $\mu$m (water vapours) and 2.4~$\mu$m \citep[from CO;][]{Blagorodnova2021}.
We performed a Planckian fit to the reddening-corrected SED, which provides an excellent match up to the red tail with a BB that has the following parameters: $T_{BB}=3980\pm60$~K, $R_{BB}=2250\pm60$~\Rsun, and $L_{BB}=(4.35\pm0.25)\times10^{39}$ erg~s$^{-1}$. These parameters are consistent with those obtained from the optical filters at similar phases.
No IR excess is detected up to 5 $\mu$m, as a single BB is sufficient to fit the data even at the reddest wavelengths.
Therefore, either dust is not present and has yet to form, or there is pre-existing dust, much colder than 600~K, otherwise its emission would have been visible as an IR excess at such long wavelengths.

\section{Spectroscopic evolution}\label{spectroscopy}
\begin{figure}
\includegraphics[width=1\columnwidth]{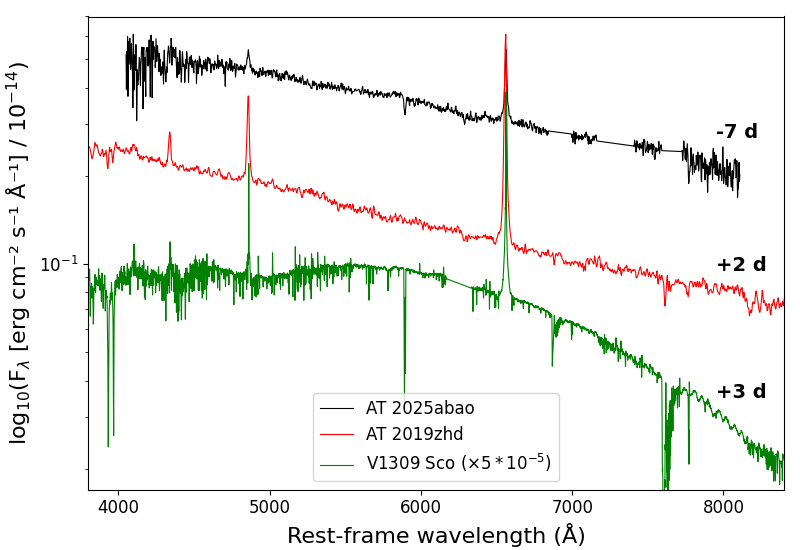}
\caption{Comparison of the earliest spectrum of \obj\ to the early spectra of the comparison objects AT~2019zhd and V1309~Sco \citep{Mason2010A&A...516A.108M}. All objects show a hot, blue continuum with narrow emissions from \Ha\ and \Hb\ lines. At this early phase, V1309~Sco already presents signs of metal absorptions.
}
\label{fig:early_spectra}
\end{figure}

The transient \obj\ was classified as an LRN on 2025 October 29, ten days after the official discovery date and at a phase of $-8.7$~d \citep{Taguchi2025ATel17468....1T}. 
However, earlier spectra from amateur astronomers are also available. The earliest, taken on 2025 October 27 (phase $-10.4$~d), was presented by \cite{Skopal2025ATel17476....1S}. 
One day later, another low-resolution spectrum was obtained by C. Balcon (phase $-9.4$~d), and additional spectra were obtained with a nearly daily cadence (at phases $-6.6$, $-5.3$, and $-4.4$~d) by \citet{Taguchi2025ATel17468....1T} and \citet{Skopal2025ATel17476....1S}.
These early spectra exhibit a blue continuum ($T_{\rm BB}=7100$ K) with superimposed narrow emission lines of the Balmer series, a feature typical of LRNe during the rise to maximum light \citep{Pastorello2019review}.
The classification spectrum of \citet{Taguchi2025ATel17468....1T}, retrieved from the TNS, is shown in Fig.~\ref{fig:early_spectra}, where it is compared with spectra of LRNe AT~2019zhd and V1309 Sco at the maximum light to emphasise their overall similarity. Nonetheless, the spectral evolution of V1309~Sco is more rapid, and metal lines are already visible in absorption at these early epochs.

\begin{figure*}\centering
\includegraphics[width=1.54\columnwidth]{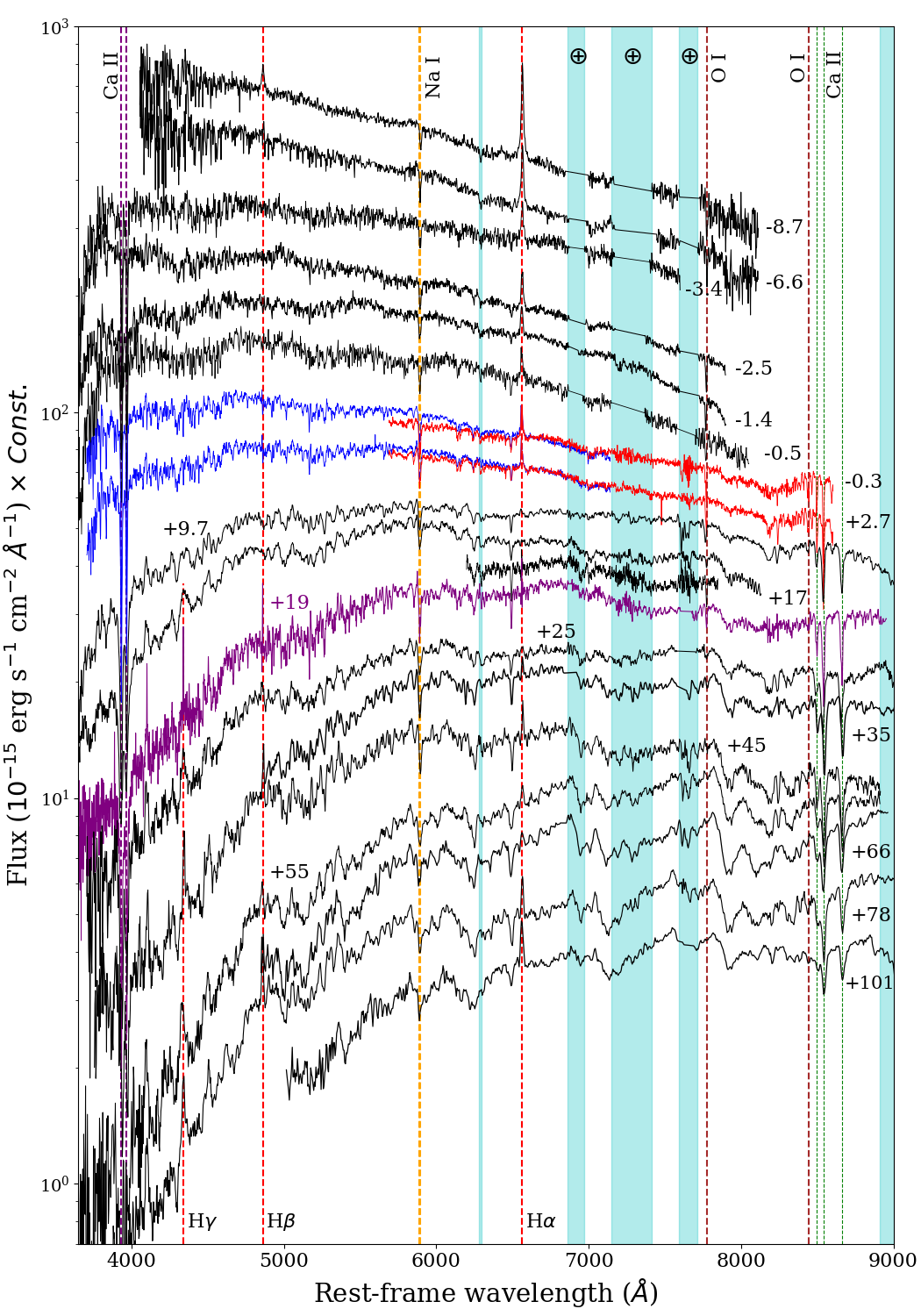}\includegraphics[width=0.47\columnwidth]{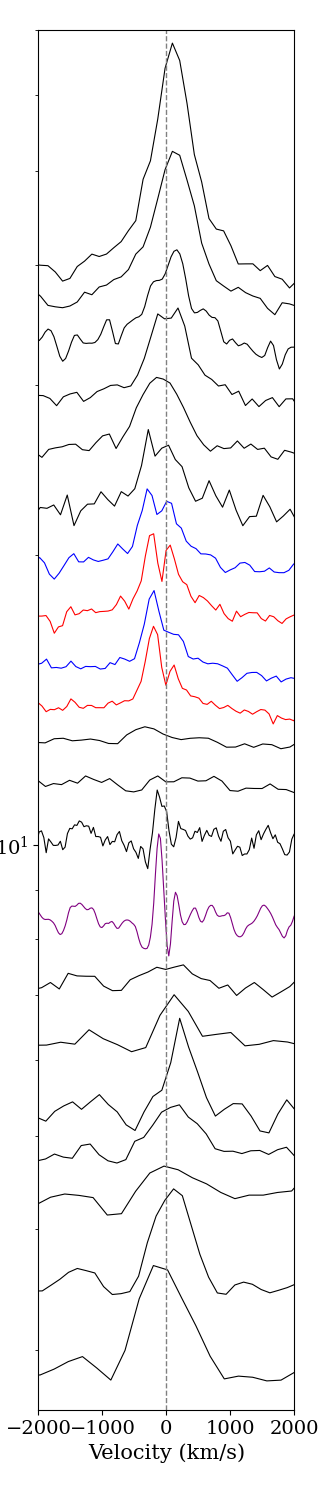}
\caption{Left: Spectral time series of \obj\ and basic line identification. The spectra are corrected for blueshift and total reddening and are plotted in logarithmic scale.
The mid-resolution GTC+OSIRIS spectrum is highlighted in purple, while the NOT+ALFOSC gr7 and gr8 spectra, taken on the same night, are shown in blue and red, respectively. Phases in days are reported on the right side of each spectrum.
The regions contaminated by telluric absorptions are identified with cyan bands.
Right: Evolution of the \Ha\ profile in the velocity space.
}
\label{fig:spectra}
\end{figure*}

We conducted a spectroscopic follow-up of \obj, during which we collected 21 optical spectra spanning more than three months of evolution. The log of spectroscopic observations is given in Table \ref{tab:spectra}, and the spectral time series is shown in Fig. \ref{fig:spectra}, left panel. 
The spectra from the 1.22m Galileo telescope equipped with a Boller \& Chivens (B\&C) spectrograph and the 1.82m Copernico telescope with an Echelle spectrograph were reduced with a custom-made, \texttt{IRAF}-based pipeline developed by A. Siviero, while the spectra from the 2.56m Nordic Optical Telescope (NOT) with ALFOSC, the 1.82m Copernico telescope with AFOSC, and the 10.4m Gran Telescopio Canarias (GTC) plus OSIRIS were reduced using the \textsc{foscgui}\footnote{{\sc Foscgui} is a graphic user interface aimed at extracting SN spectroscopy and photometry obtained with FOSC-like instruments. It was developed by E. Cappellaro. A package description can be found at \url{http://sngroup.oapd.inaf.it/foscgui.html}.} pipeline. Finally, the spectra of the 3.58m Telescopio Nazionale Galileo (TNG) plus DOLORES spectra were hand-reduced using routine \texttt{IRAF} procedures. 

\subsection{Main spectral evolution}
\begin{figure*}\centering
\includegraphics[width=1\columnwidth]{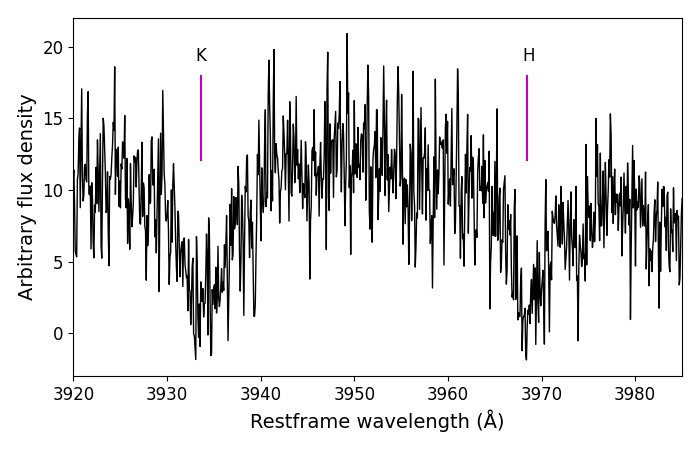}\includegraphics[width=1\columnwidth]{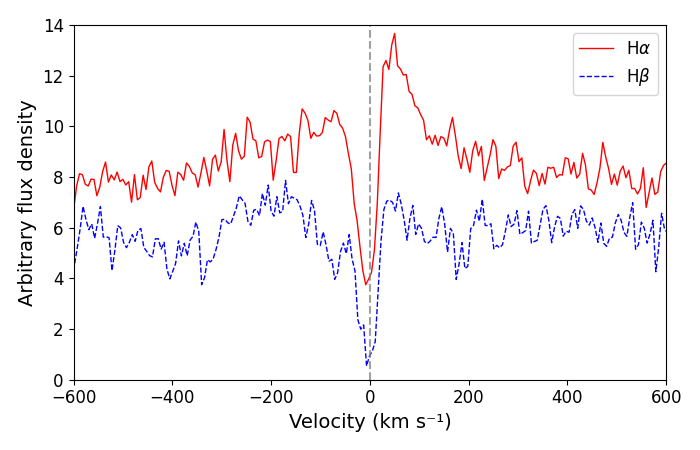}
\caption{Zoomed-in view of some spectral lines visible in the $-3.4$~d Echelle spectrum. Left: \Caii\ H\&K. Right: \Ha\ and \Hb, overlapped in the velocity space, with a narrow blueshifted absorption clearly visible.
}
\label{fig:echelle}
\end{figure*}

On 2025 November 3 (phase $=-3.4$ d), we obtained a high-resolution ($R\sim20,000$) Echelle spectrum with the Copernico 1.82-m telescope, in which we can identify specific spectral features. The main spectral features identified in this spectrum are highlighted in Fig. \ref{fig:echelle}.
\Ha\ and \Hb\ have a narrow absorption profile with a width of 45 \kms, on top of a much broader emission component ($\sim$450 \kms, Fig. \ref{fig:echelle}, right panel). 
Along with the H lines, we clearly identify prominent \Nai\ doublet features, and all the components well discerned. However, while the absorption lines of the \Nai\ doublet at redshift zero are unresolved and hence can be attributed to material along the line of sight within the MW, the \Nai\ doublet lines at the redshift of M~31 are resolved and have a FWHM of around 20-25 \kms\ (see Fig. \ref{fig:Na_echelle} in Appendix \ref{Appendix_B}, top panel). This FWHM is similar to those exhibited by other metal lines visible in the $-3.4$~d spectrum (see below); hence, we consider these \Nai\ features as intrinsic to the transient. A prominent absorption feature is observed at 5890.9~\AA, which is unlikely to be an additional Na~I component, and whose nature is discussed in Appendix~\ref{Appendix_B}. 
Lines of the Mg I triplet at 5160--5180~\AA\ are also detected, with a FWHM similar to that of the \Nai\ lines. More metal lines are detected bluewards of 5200~\AA, including from \Feii, which also have a FWHM of 25~\kms. Finally, broad \Caii\ H\&K absorption lines are visible in the UV region, with a much higher FWHM of $\sim500-600$~\kms\ (Fig.~\ref{fig:echelle}, left panel). The broadness of the \Caii\ H\&K lines is due to the elevated opacity of those transition lines, as effects of saturation are noticeable and possibly also blending with other metal lines. 
However, it is still possible to have a fast-moving outflow, produced after the merger of the stars' cores, perpendicular to the orbital plane (see \citealt{Metzger2017MNRAS.471.3200M, Gagnier2025A&A...697A..68G}) and with a higher velocity than the material along the equatorial disc.

\begin{figure}\centering
\includegraphics[width=1\columnwidth]{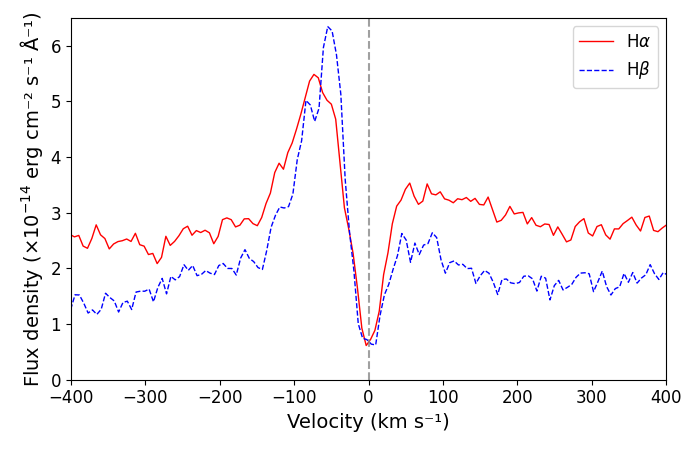}
\caption{Zoomed-in view of the \Ha\ and \Hb\ lines, in velocity space, in the $+34$~days Echelle spectrum.
}
\label{fig:echelle2}
\end{figure}

In the spectra taken at about the $g$-band maximum light, together with a double-horned \Ha\ line, a faint \hei\ 5876 is the only other transition seen in emission, on top of a continuum peaking at 4800--4900 \AA\ (indicating $T_{\rm BB}\sim$6000 K, which is consistent with the SED inferred in Sect. \ref{Sect:SED}). The most notable features are the deep and strong absorptions of the \Caii\ H\&K UV doublet, \Nai, and a forest of metal lines. In particular, the \Feii\ multiplet 42 ($\lambda\lambda\lambda$ 4924, 5018, 5169) transitions are clearly detectable. 

\begin{figure*}\centering
\sidecaption
\includegraphics[width=1.4\columnwidth]{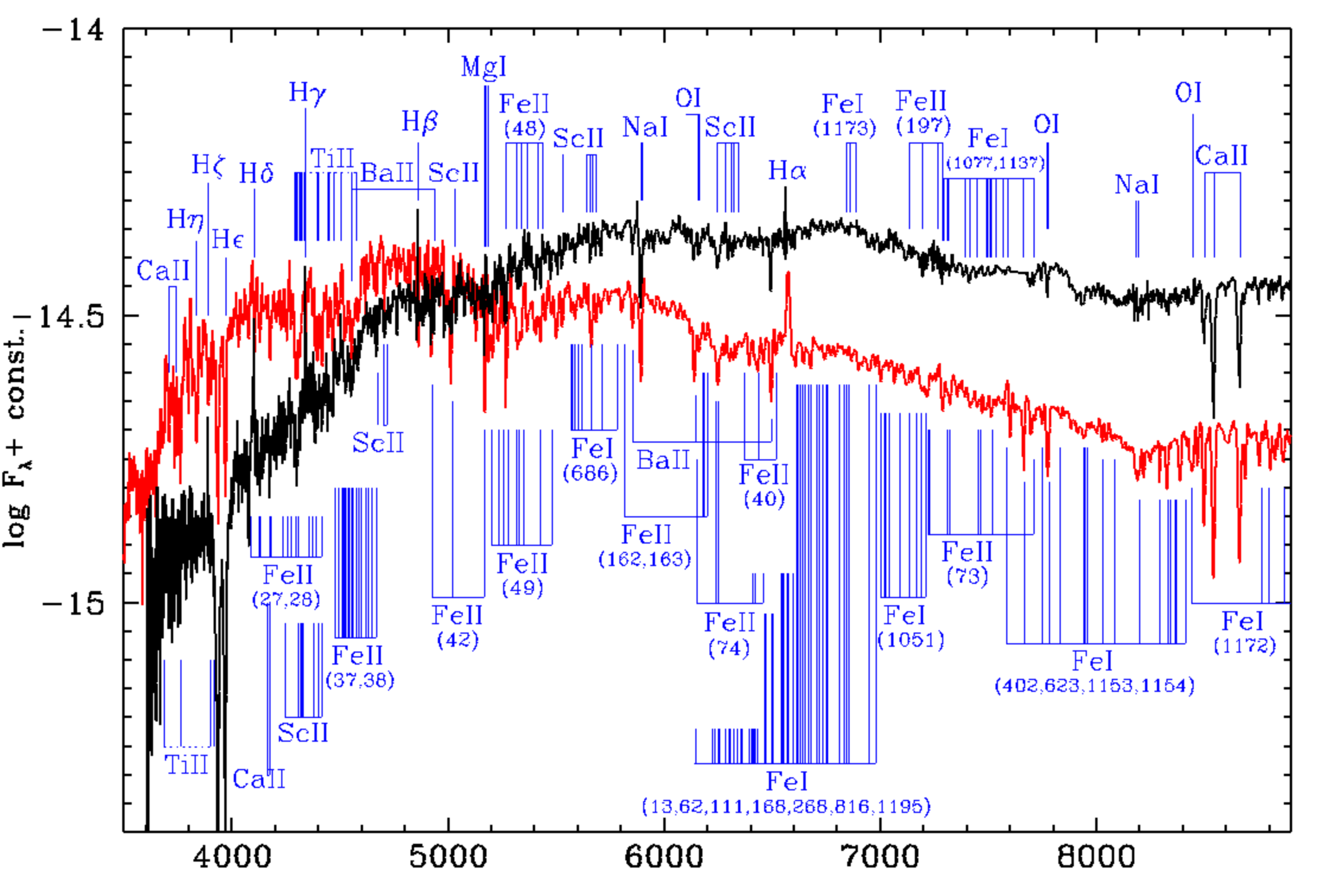}
\caption{Line identification on the mid-resolution spectrum of \obj\ taken with GTC+OSIRIS at $+18.5$ d (black), compared with a late time ($+103$ d) spectrum of LRN AT 2011kp \citep[red line; from][]{Pastorello2019review}. Transitions from multiplets of neutral and singly ionised metals are marked.
}
\label{fig:lines}
\end{figure*}

By the $+2.7$~d spectrum, the continuum temperature has already declined to 5200 K (with the spectral continuum emission peaking at around 5600 \AA). The gas cooling is confirmed by the disappearance of the He~I 6678 line. Otherwise, the spectrum did not change significantly, with the most prominent metal transitions still clearly present.
In the $+9.7$~d spectrum, the continuum peaks at around 5900 \AA, indicating a temperature of only $\sim$4900~K. Emission lines, including \Ha, have completely disappeared, and visible features are only in absorption. Multiple metal lines are identified in the bluer region, together with the usual \Caii\ (both H\&K and IR triplet) and \Nai.
Little has changed in the spectrum at $+16.7$~d. A broad basin is visible between 4800 and 5300 \AA, a feature commonly seen in the spectra of Sun-like stars, and attributed to the absorption due to the CH radical molecule.

In the $+24.7$ d spectrum, as already in the $+18.5$ d spectrum, the continuum peak has shifted to around 6800 \AA\ ($T\sim$4250 K), maintaining a steep decline towards the UV likely due to line blanketing, while being flat at redder wavelengths.
The strongest features are now the absorptions from the \Caii\ NIR triplet, though the \Caii\ H\&K absorption lines are also recognisable, together with the depression due to CH, and even more numerous metal lines. Balmer lines are now weak or absent.

On December 10 (phase $=+33.6$~d), we took another Echelle spectrum, which reveals a peculiarity: the \Ha\ line presents a narrow absorption redwards of the centre of the emission. The velocity of the minimum of this feature relative to the emission peak is $\sim +70$ \kms.
The same structure is also detected in the \Hb\ line (see Fig.~\ref{fig:echelle2}, where its profile is overlapped with that of \Ha). The feature looks similar to a counter P~Cygni profile. An inverse P~Cygni profile has been observed before in LRN spectra (see AT~2011kp; \citealt{Pastorello2019review}) and in another IGT, AT~2008jd \citep{Berger2009ApJ...699.1850B}. In the latter case, it was interpreted as infalling material from a disc on the orbital plane towards the central system through the second Lagrangian point \citep[see also][]{Pejcha2016bMNRAS.461.2527P}.
However, in the case of \obj\, the emission and absorption components might not be correlated and be dynamically distinct \citep{Mason2022A&A...664A..12M}.
This second Echelle spectrum also shows a crowded line population in the 5850-5900 \AA~region: a blueshifted \hei\ $\lambda$5876 emission line, two \Nai\ doublets (one due to line-of-sight material in the MW, and the other intrinsic to \obj), and two more lines at $\lambda$5890.9 and $\lambda$5896.9, also discussed in Appendix~\ref{Appendix_B}.

In the $+45$~d spectrum of \obj, the first signs of molecular absorption bands are seen: TiO features start to appear between 6000 and 8000 \AA, in the form of two sawtooth drops in the continuum, which are sharp on the blue edge and then gradually rise towards the red edge. 
This trend of molecule formation is confirmed in the $+66$~d spectrum. Molecular bands from VO and TiO are clearly present in the regions 6100--6700 \AA\ and 7100--7800 \AA, together with a rise at the start of the NIR, in the 8400--8800~\AA\ range, also compatible with a molecular band. The continuum peaks at around 7800~\AA\ ($T\sim$3700 K), with a temperature trend consistent with that expected for a moderately faint LRN.
Balmer lines appear to be strong again in the $+78$ and  $+101$~d spectra, while the continuum is now red and its temperature has decreased to 3600 and 3500~K, respectively.

\subsection{Spectral line identification}
On November 25 ($+18.5$ d), four medium-resolution ($R\sim2,100$) spectra of \obj\ were obtained with 10.4m GTC+OSIRIS and the R2500-class grisms, covering wavelengths from the $U$ to the $I$ bands. In these spectra, we identified the most prominent metal lines, highlighting that these features are real and not noise patterns, since for each grism we took four exposures in sequence and checked whether a line was present in all individual spectra. 

In the GTC spectrum, \Ha\ shows a narrow P-Cygni profile with a dominant absorption with a minimum blueshifted by $-300$ \kms\ (although the blue wing extends down to $-450$ \kms), and a weaker and narrow emission component. However, on the red side, a sharper counter P~Cygni is also present. Other Balmer lines are detected (\Hb, \Hg, \Hd, and possibly \He), dominated by the emission component.
\Caii\ H\&K are relatively broad (FWHM$=800-900$ \kms) and in pure absorption. Other notable absorption features are the \Caii\ NIR triplet, and the \Nai\ doublet. All other features, while present and identified, are less strong.

In Fig. \ref{fig:lines}, we compare the full GTC+OSIRIS mid-resolution spectrum of \obj\, with that of AT 2011kp during late phases ($+103$ d) from \cite{Pastorello2019review}, the authors of which also performed a similar line identification in a mid-resolution spectrum.
The spectrum of \obj\ was taken 20 days after maximum, while that of AT 2011kp was obtained after more than three months. While the spectral continuum shows some differences, the most important metal lines are identified in both spectra. This is an indication of a much faster evolution of \obj. 
Thus, a similar set of lines is identified and seen in both spectra: multiplets from \Feii, \Fei, \ion{Mg}{i}, \ion{Sc}{ii}, \ion{Ba}{ii} (the strongest line being $\lambda$6497), and \ion{Ti}{ii} are retrieved. Single lines from H, \ion{O}{i}, \Caii, and Na I are also found. In the red extreme, in a region not strongly contaminated by telluric absorption bands, we also identify the Na I $\lambda\lambda$8183,8195 doublet, \ion{O}{i} $\lambda$7774 and $\lambda$8446.
Many lines, not marked in Fig. \ref{fig:lines}, can be attributed to a multitude of transitions from \ion{Ti}{ii}, as also suggested by \cite{Pastorello2019review}.

\section{Discussion and conclusion}\label{discussion}

\obj\ is the most recent example of an LRN in M~31, and the fourth in the last 40 years in this galaxy, with an observed LRN-event rate of $\approx0.1\ \rm yr^{-1}$. All of them have peak absolute magnitudes in the $-8.5$ to $-10$ mag range. Brighter objects would have been easily detected, while fainter ones could be lurking among the Novae population.
For comparison, only one object (V838 Mon) was discovered with this luminosity range in a similar lapse of time in the MW.
Based on the estimations of \cite{Kochanek2014MNRAS.443.1319K}, we calculated an event rate of just 0.04 yr$^{-1}$ within the same magnitude range (0.07 yr$^{-1}$ of events brighter than $-8.5$ mag minus 0.03 yr$^{-1}$ brighter than $-10$ mag).
Therefore, it seems that this kind of transient is more than twice as common in the Andromeda galaxy than in the MW. 
The stellar mass of M~31 has been estimated to be $1-1.5\times10^{11}$ \Msun\ \citep{Barmby2006ApJ...650L..45B, Tamm2012A&A...546A...4T, Kafle2018MNRAS.475.4043K}, while the equivalent for the MW is $5-6\times10^{10}$ \Msun\ \citep{Licquia2015ApJ...806...96L, Bland-Hawthorn2016ARA&A..54..529B}. Hence, the Andromeda galaxy is $\sim2.3$ times more massive in terms of stellar content and, likely, in its number of stars. If we scale up the Galactic LRN rate in the $-8.5$ to $-10$ mag range from \cite{Kochanek2014MNRAS.443.1319K} by this ratio, we find a rate that is very similar to the observed one.
The LRN population is dominated by faint events, with 86 \% of them having $M_V$ between $-3$ and $-8.5$ mag, while those in the $-8.5$ to $-10$ mag range make up only 8\%, and those more luminous than $-10$ mag just 6\% (however, \citealt{Karambelkar2023ApJ...948..137K} predicted an even lower rate for these brighter events).
Therefore, while we expect more than one LRN per year in M~31 ($0.5\ \rm yr^{-1}$ for $M_V>-3$ mag according to \citealt{Kochanek2014MNRAS.443.1319K}, times $2.3$ equals $\approx 1.15\ \rm yr^{-1}$), the rate of $M_V>-10$ events should only be $\sim0.07\ \rm yr^{-1}$ (or less according to \citealt{Karambelkar2023ApJ...948..137K}), making them particularly rare.

Usually, the progenitors of LRNe are found to be yellow stars, often supergiants in the cases of the most luminous events \citep[][]{Mauerhan2015MNRAS.447.1922M, Blagorodnova2017, Pastorello2021_20hat, Blagorodnova2021, Cai2022A&A...667A...4C}. 
However, the association of \obj\ with \WNT, a slowly variable AGB star \citepalias{Karambelkar2025ApJ...993..109K}, makes it an exception. \citetalias{Karambelkar2025ApJ...993..109K} interpreted this object as a common-envelope event and predicted a merging event as the outcome.
AGB stars are so inflated that the outer layers of the atmosphere can be considered a hot vacuum. Therefore, the companion star --especially if the mass ratio is particularly large-- could pass through it, remaining engulfed within the envelope without significantly changing the global colour of the system, which would remain red, as observed by \citetalias{Karambelkar2025ApJ...993..109K}.

\begin{figure}\centering
\includegraphics[width=\columnwidth]{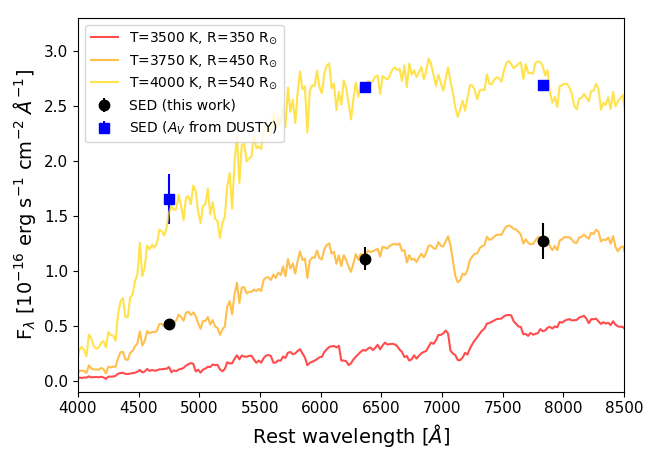}
\caption{Spectral energy distribution of the precursor of \obj, corrected for extinction in two different methods, compared to an ATLAS9 stellar atmospheric model. A~good match is found with a model with $T_s=4000$ K and $R_s=540$~\Rsun.
The model corresponding to the quiescent progenitor as determined by \citetalias{Karambelkar2025ApJ...993..109K} is shown in red.}
\label{fig:Kurucz}
\end{figure}

We constructed the pre-outburst SED of \obj\, at the time of the last local peak in the light curve (at $-$333 d), before the optically thick phase. We only used the $gri$-band data from ZTF, and since there was no contemporary $i$-band imaging, we assumed a constant $r-i$ colour from the only pre-outburst $i$-band public observation by ZTF (at $-$411 d, $r-i=+0.62$ mag).
Subsequently, we corrected the SED for the same Galactic extinction adopted by \citetalias{Karambelkar2025ApJ...993..109K} and accounting for their dust parameters obtained with the radiative transfer code \texttt{DUSTY} \citep{Ivezic1997MNRAS.287..799I} for the quiescent progenitor (i.e. $\tau_V=2.7$, $T_s=3500$~K, $T_d=1450$~K, $Y=2$, $\rho \propto r^{-2}$, silicate composition). These parameters, according to our \texttt{DUSTY} simulation, result in an additional circumstellar extinction of $A_V=1.025$ mag.
We compared this corrected SED with the ATLAS9 stellar atmospheric models by \cite{Castelli2003IAUS..210P.A20C}\footnote{Retrieved here: \url{https://archive.stsci.edu/hlsps/reference-atlases/cdbs/grid/ck04models/}}. We adopted the models with $Z/Z_{\odot}=1$ and log$(g)=0.0$. 
We find a good match with a model with a photospheric temperature of 4000 K, and a stellar radius of about 540 \Rsun.
The match of the ATLAS9 models to the precursor SED is shown in Fig.~\ref{fig:Kurucz}.
These parameters are consistent with those of an orange supergiant, but also with an expanded AGB star, though hotter and more luminous (log$(L_*/L_{\odot}) = 4.83$ vs. 4.20) than the progenitor parameters obtained by \citetalias{Karambelkar2025ApJ...993..109K}.
If we instead adopt our distance and reddening estimates (Sect. \ref{discovery}), the corrected SED is fitted with a model that has $T_s=3750$ K, a radius of $\sim450$ \Rsun, and a corresponding log$(L_*/L_{\odot}) = 4.57$, which is still hotter and more luminous than the quiescent progenitor.

\begin{figure}\centering
\includegraphics[width=\columnwidth]{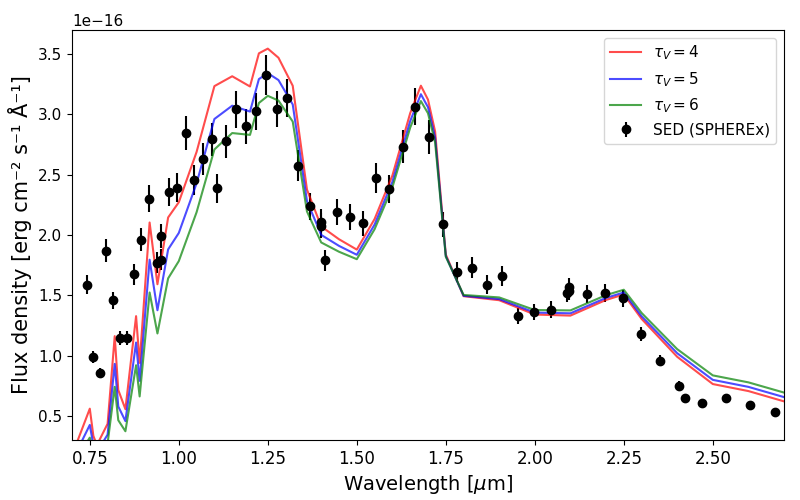}
\caption{Matches of the spectral energy distribution of the precursor of \obj\ observed by SPHEREx with
three \texttt{DUSTY} models. They are constructed using a PHOENIX atmospheric model of $T_{\rm eff}=3100$~K extinguished by a dusty shell with silicate composition, $T_d=1450$~K, and $\tau_V=4,5,6$.}
\label{fig:SPHEREX_PHOENIX_DUSTY}
\end{figure}

We also used \texttt{DUSTY} to model the SED of the precursor of \obj\ observed by SPHEREx a few months before the outburst, at a phase close to the optical minimum.
Since the broad molecular absorptions at 1.4-1.5 and 1.8-2.2 $\mu$m are large deviations from a BB, we opted for an atmospheric model of a cool star as the radiation source. We adopted the PHOENIX models\footnote{Retrieved here: \url{https://pollux.oreme.org/explore/BT-Dusty/}} \citep{Husser2013A&A...553A...6H} with effective temperatures, $T_{\rm eff}$, between 2500 and 3500 K, $[Fe/H]=0.0$, and log$(g)=1.0$. 
Good matches are found for the model with $T_{\rm eff}=3100$ K and $\tau_V=5\pm1$, which is higher than that found for the quiescent progenitor.
The matches of the PHOENIX+\texttt{DUSTY} models to the SED from SPHEREx are shown in Fig.~\ref{fig:SPHEREX_PHOENIX_DUSTY}.
The inner radius of the shell of dust is $\sim1180\pm50$ \Rsun\ ($\sim5.5$ AU), while the radius of the source is 350 \Rsun, the same as the quiescent progenitor.
However, the dust mass we obtain, following the procedures of \cite{Reguitti2026}, is only $\approx1.5\times10^{-8}$ \Msun; that is, one order of magnitude lower than \citetalias{Karambelkar2025ApJ...993..109K}. This is due to the inner radius of the dust shell being much smaller.

\citetalias{Karambelkar2025ApJ...993..109K} estimated a mass of the progenitor of $7\pm2$ \Msun\ based on evolutionary paths from MESA isochrones and stellar tracks. Using the empirical relation between absolute magnitude and the mass of the progenitor by \cite{Cai2022A&A...667A...4C} for an object with $M_V=-9.74\pm0.1$ mag during the plateau phase\footnote{Obviously, \obj\ does not exhibit a genuine secondary red peak, but instead a long plateau. However, the relation of \cite{Cai2022A&A...667A...4C} can still be used, considering the $V$-band absolute magnitude at the end of the plateau.
Consequently, in Appendix~\ref{Appendix_A}, the \cite{Cai2022A&A...667A...4C} relation has been updated using also the \obj\ data.}, 
we obtained $7.5^{+6.1}_{-3.4}$ \Msun, which is in excellent agreement with their estimation.
Based on the peak luminosity and similarities in the spectral evolution, we can also state that the progenitor mass of \obj\ must be similar to that of V838~Mon, which was $8\pm3$ \Msun\ \citep{Pastorello2023, Tylenda2005_V838} and slightly more massive than that of AT~2019zhd (3.4 \Msun\ for \citealt{Pastorello2021_19zhd}; 6 \Msun\ for \citealt{Chen2024ApJ...963L..35C}).

To infer additional physical parameters, we considered the analytical expressions derived by \cite{Matsumoto2022ApJ...938....5M}. In particular, their Equation 16 allowed us to estimate the mass of the ejected material, $M_{\rm ej}$, by measuring the duration of the plateau, $t_{\rm pl}$, and the bulk velocity of the ejecta, $\overline{v}_{\rm E}$. For the plateau length, we adopted a value of $t_{\rm pl}=80$ days, as it started in the $V$ band one week before maximum light and ended at $+72$~d. For the ejecta velocity, we adopted the FWHM of the \Ha\ line estimated during the plateau (as in \citealt{Cai2022A&A...667A...4C}), precisely in the $+35$ and $+45$ d spectra; hence, $\overline{v}_{\rm E}\simeq450$ \kms. The same velocity was obtained by measuring the FWHM of the broad component in the \Ha\ emission line in the first Echelle spectrum, and from the blue wing of the P~Cygni profile of the same line (and \Hb) in the GTC spectrum.
Under these assumptions, we calculated an ejecta mass of $\sim$0.63~\Msun. This $M_{\rm ej}$ value roughly coincides with the one found by \cite{Wu2026}, who modelled our bolometric light curve. 
The associated kinetic energy is $1.6\times10^{48}$ erg; as a comparison, the total kinetic+radiated energy emitted by V838 Mon was $(0.3-1)\times10^{48}$ erg \citep{Tylenda2006A&A...451..223T}. The collision of the two stellar cores deposits sufficient energy \citep[see also][]{Bally2005AJ....129.2281B} in the outer layers to lift the ejecta at that encountered high speed.

\obj\ is an LRN exhibiting a plateau instead of a double-peaked light curve. Its progenitor has been established as an early AGB \citepalias{Karambelkar2025ApJ...993..109K}, that is, an expanded star with an H-rich envelope. The photometric behaviour of \obj, similar to that of a Type IIP SN, can be explained with the recombination of a large quantity of H, previously ionised by the shocks following the merger of the two stars \citep{MacLeod2017ApJ...835..282M}.
Therefore, we propose that the observed dichotomy in LRNe --with some showing two clear light-curve peaks (the first blue and the second red) and others with a plateau instead-- is linked to the extent of the common envelope and the amount of H. 
Hence, systems with compact progenitors or a low H content would be responsible for double-peaked LRNe, whilst events with extended progenitors and massive outer H layers would produce LRNe with flat light curves (see also \citealt{Howitt2020MNRAS.492.3229H, Twum2026arXiv260210211T}).
The total mass of the system instead influences the luminosity of the event (as established by \citealt{Kochanek2014MNRAS.443.1319K, Blagorodnova2021, Cai2022A&A...667A...4C}).
Another possibility is that the plateau luminosity is maintained long by shocks produced by the collision of shells, together with the H recombination \citep{Matsumoto2022ApJ...938....5M, Kirilov2025ApJ...994L..41K}.

\begin{figure}\centering
\includegraphics[width=\columnwidth]{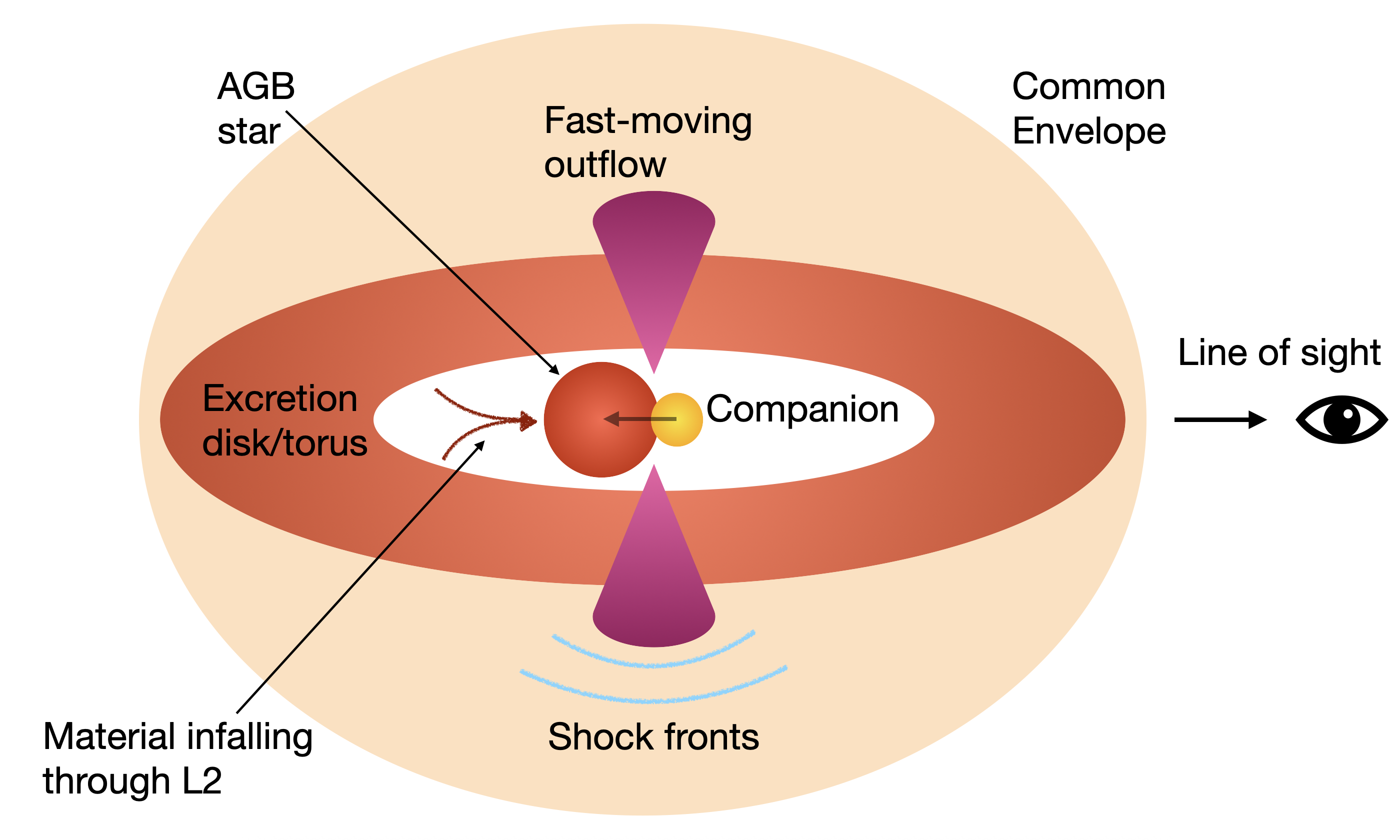}
\caption{Schematic sketch presenting the proposed scenario. At the centre, the companion star is merging with the primary AGB star, within a common envelope. The cores of the two stars collide, and the kinetic energy is transferred to the fast-moving outflow, ejected along the polar direction. Shock fronts can be produced when the outflow hits the common envelope. On the orbital plane, the system is surrounded by a disc- or torus-shape structure, made of slow-moving material, some of which could also infall towards the system via the L2 point. The entire system is observed almost edge-on.}
\label{fig:sketch}
\end{figure}

The double-peaked \Ha\ profile in the spectra at maximum light (Fig. \ref{fig:spectra}, right panel), with the red `horn' fainter than the blue one, can be explained by an excretion circumbinary disc \citep[which may form if not all of the envelope is ejected;][]{Ivanova2013A&ARv..21...59I} or a torus-shaped structure surrounds the system.
We propose a scenario in which this disc or torus would be observed at a nearly edge-on viewing angle and would obscure the receding side of the CSM, making the redshifted portion of it more attenuated and thus fainter \citep[see also][]{Kirilov2025ApJ...994L..41K}.
A fast outflow ($\sim$450 \kms), far from being spherically symmetric (though it is still possible it could be axisymmetrical), is instead responsible for the broad emission component underneath the double-peaked profile. 
This model is similar to that presented by \cite{Kashi2017MNRAS.467.3299K}.
We present a sketch of this scenario in Fig.~\ref{fig:sketch}.

\begin{table}[b]\centering
\caption{Main parameters of the proposed scenario observed.}
\label{tab:parameters}
\begin{tabular}{lc}
\hline
Mass of the primary star ($M_*$) & $7\pm2$ \Msun \\
Radius of the primary star ($R_*$) & $\sim350$ \Rsun \\
Temperature of the primary star ($T_*$) & 3100 K \\
Viewing angle & $\sim90^{\circ}$ \\
Ejecta mass ($M_{ej}$) & $\sim0.63$ \Msun \\
Fast outflow velocity ($v_{ej}$) & $\sim450$ \kms \\
Kinetic energy ($E_k$) & $\sim1.6\times10^{48}$ erg \\
Velocity of the material on the orbital plane & $25-70$ \kms \\
Total radiated energy & $>3.1\times10^{46}$ erg \\
\hline
\end{tabular}
\end{table}

In conclusion, it will be crucial to observe \obj\ years after the event with space telescopes, especially in the infrared domain (as done by \citealt{Karambelkar2026ApJ...999...16K}, or with the forthcoming Roman, \citealp{Akeson2019arXiv190205569A}), to see what kind of outcome remains and if the system returns to a state similar to the pre-LRN one, with an inflated red star. We note that a red (super)giant survivor was observed after the coalescence of a number of LRNe \citep{Steinmetz2025A&A...699A.316S, Reguitti2026}.
Future observations will also constrain the variability of the post-merger remnant, which may confirm whether the stellar components of the system had completely coalesced. In fact, very-late-time observations of V838 Mon by \cite{Goranskij2020AstBu..75..325G} revealed a slow brightening in the optical, which raises doubts as to whether a complete merger really occurred in that system.

\section*{Data availability}
The observed calibrated magnitudes are tabulated in the photometry table, which is only available in electronic form at the CDS via anonymous ftp to cdsarc.u-strasbg.fr (130.79.128.5) or via http://cdsweb.u-strasbg.fr/cgi-bin/qcat?J/A+A/.
All the spectra are released on the \textsc{WISeREP} interface (https://www.wiserep.org/).

\begin{acknowledgements}
\begin{small}
We thank the anonymous referee for the useful comments and suggestions to improve the manuscript.
AR, GV, YZC acknowledge financial support from the SOXS project (PI S. Campana). AR, AP, GV, NER acknowledge financial support from the PRIN-INAF 2022 "Shedding light on the nature of gap transients: from the observations to the models".
NER acknowledges support from the Spanish Ministerio de Ciencia e Innovaci\'on and the Agencia Estatal de Investigaci\'on 10.13039/501100011033 under the program Unidad de Excelencia Mar\'ia de Maeztu CEX2020-001058-M.
TMR acknowledges support from the Research Council of Finland project 350458 and the Cosmic Dawn Center, funded by the Danish National Research Foundation under grant DNRF140. 
MDS is funded by the Independent Research Fund Denmark (grant nr. 10.46540/2032-00022B).
YZC is supported by the National Natural Science Foundation of China (Grant N. 12303054), the National Key Research and Development Program of China (N.~2024YFA1611603), the Yunnan Fundamental Research Projects (N.~202501AS070078), and the International Centre of Supernovae, Yunnan Key Laboratory (N.~202302AN360001).
Based on observations collected at Copernico and Schmidt telescopes (Asiago Ekar, Italy) of the INAF-Osservatorio Astronomico di Padova, and at the Galileo telescope of the Padova Univesity (Asiago Pennar, Italy).
AR thanks Alessandro Fabris and Virginia Albanese for having participated to the observations with AFOSC.
Support to ATLAS was provided by NASA grant NN12AR55G.
Based on observations made with the NOT, owned and operated jointly by Aarhus University, Turku University, Oslo University, the University of Iceland and Stockholm University, under program 72-305 (PI Reguitti), and via the NUTS2 collaboration which is supported in part by the Instrument Centre for Danish Astrophysics, and the Finnish Centre for Astronomy with ESO via Academy of Finland grant nr. 306531.
Based on observations made with the GTC. 
Based in part on observations made at the TNG, operated by INAF, under program A50TAC\_41 (PI Valerin). 
All three telescopes are installed at the Spanish Observatorio del Roque de los Muchachos of the Instituto de Astrofísica de Canarias, on the island of La Palma.
We acknowledge the use of data from the Swift data archive.
FR is grateful to iTelescope.Net for providing observing
time to use their remote telescopes and to the AAVSO who made it possible to use the AAVSOnet telescopes.
SPHEREx is a joint project of the Jet Propulsion Laboratory and the California Institute of Technology, and is funded by the National Aeronautics and Space Administration.
\end{small}
\end{acknowledgements}

\bibliographystyle{aa} 
\bibliography{bib,fate}

\begin{appendix} 

\section{Lines in the \Nai\ region in the Echelle spectra}\label{Appendix_B}
\begin{figure}[h]
\centering
\includegraphics[width=1\columnwidth]{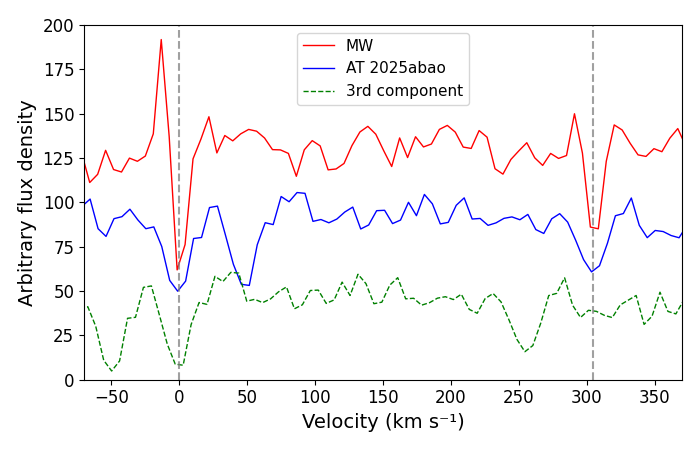}
\includegraphics[width=1\columnwidth]{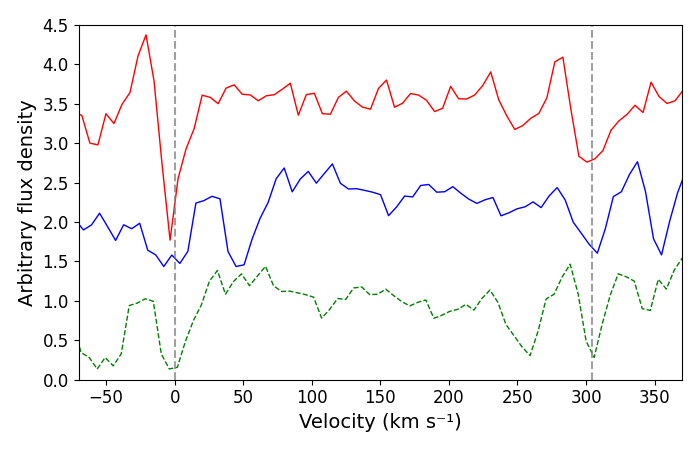}
\caption{Zoomed-in view of the \Nai\ spectral region in the Echelle spectra. The \Nai\ doublets of the MW (red solid line) and that of the LRN (blue solid line) are plotted in the velocity space relative to the $\lambda$5890 line (\Nai2). 
Top: $-3.4$ d spectrum. The feature observed at $\lambda$5890.9, indicated as third component (green dashed line), is also shown at zero velocity. Adopting a tentative classification as an additional \Nai2 component redshifted by $\sim$50 \kms\ (e.g., from a gas cloud along the line of sight), we note however that the corresponding \Nai1 line is missing in this early spectrum. 
Bottom: $+33.6$ d spectrum. While the $\lambda$5890.9 feature is still visible, a new line appears at $\lambda$5896.9 at the expected wavelength of \Nai1, that we preferentially identify as an \Fei\ or a \ion{V}{ii} transition (see text).
}
\label{fig:Na_echelle}
\end{figure}

In the first Echelle spectrum of \obj\ at $-3.4$ d (Fig.~\ref{fig:Na_echelle}, top panel), a prominent absorption feature with a profile and intensity similar to the \Nai2 line due to the transient is visible at 5890.9 \AA\ (in the host galaxy rest frame). This has been tentatively identified as the transition \Nai2 from a third Na I doublet, redshifted by $\sim$50 \kms\ with respect to that of the LRN. However, the corresponding \Nai1 transition, which would be located at 5896.9~\AA, is not detected.
Therefore, we believe that this feature is not an additional \Nai\ component, but most likely an absorption metal line. One possibility (though uncertain) is \Feii\ $\lambda$5891.48 (multiplet 211), which is consistent with the detection of other \Feii\ lines in the spectrum. However, no other \Feii\ lines from the same ionisation stage has been found.

In the second Echelle spectrum at $+33.6$ d (Fig.~\ref{fig:Na_echelle}, bottom panel), the $\lambda$5890.9 line is still present, but a new absorption is now visible at 5896.9 \AA, which is fortuitously at the expected position of the \Nai1 transition, assuming that the doublet is redshifted by 50 \kms\ with respect to the transient.
However, accounting that more absorption lines of metals in neutral state had appeared at this phase, as the continuum temperature decreased, we tentatively identify this new line as \Fei\ $\lambda$5896.74, though an identification as \ion{V}{ii} $\lambda$5897.54 is also possible, as absorption bands from the VO molecule are observed in the spectra at later phases.

\section{Updated progenitor mass -- luminosity relation}\label{Appendix_A}
Considering the progenitor of \obj\ identified in archival images before the outburst, and its mass determined independently by \citetalias{Karambelkar2025ApJ...993..109K}, we can add it to the sample of objects used by \cite{Cai2022A&A...667A...4C} to derive the empirical relation between the progenitor mass and $V$-band absolute magnitude at the second peak (or during the plateau phase) for LRNe, with the goal of updating it with the robust data inferred for \obj. 
Adopting their estimate of $7\pm2$ \Msun, and our value of $M_{V,plateau}\simeq -9.74\pm0.1$ mag, the updated mass-luminosity relation becomes:

\begin{equation}\label{eqn}
    {\rm log} (M/M_{\odot})=(-0.172\pm0.017) M_V {\rm (mag)} - (0.811\pm0.042).
\end{equation}

\onecolumn 
\section{Complementary tables}\label{Appendix_C}

\begin{table*}[h] 
\caption{Observational facilities and instrumentation used in our photometric follow-up of \obj.}
\label{tab1}
\begin{tabular}{llll}
\hline
Telescope & Location & Instrument & Filters \\
\hline
\textit{Swift} (0.3m) & Space & UVOT & $UV$ filters+$UBV$ \\
T21 (0.43m)       & Utah     & CCD      & $BVR_CI_C$ \\
ATLAS (0.50m)     & Mauna Kea& ACAM1    & $c,w,o$ \\
T11 (0.51m)       & Utah     & CCD      & $BVR_CI_C$ \\
MPO61 (0.61m)     & Texas    & CCD, CMO & $BVgriz$ \\
Schmidt (0.67m)   & Asiago   & Moravian & $uBVgri$ \\
Oschin (1.20m)    & Palomar  & ZTF      & $gr$ \\
Copernico (1.82m) & Asiago   & AFOSC    & $uBVgriz$ \\
NOT (2.56m)       & La Palma & ALFOSC   & $uBVgriz$ \\
NOT (2.56m)       & La Palma & NOTCam    & $JHKs$ \\
\hline
\end{tabular}
\end{table*}

\begin{table*}[h]
\caption{Log of the spectroscopic observations of \obj.}
\label{tab:spectra}
\begin{tabular}{llllll}
\hline
Date & MJD & Phase & Spectral & Resolution & Telescope + Instrument + Grism \\
 & & (d) & range (\AA) &  & \\
\hline
2025-10-28 & 60976.90 & $-$9.4 & 4100-8390 & $R\sim200$ & BL41 0.2m + FOSC-E5535 \\ 
2025-10-29 & 60977.61 & $-$8.7 & 4050-8100 & $R\sim500$ & Seimei 3.8m + KOOLS-IFU ($\star$) \\
2025-10-31 & 60979.72 & $-$6.6 & 4050-8100 & $R\sim500$ & Seimei 3.8m + KOOLS-IFU ($\star$) \\
2025-11-03 & 60982.86 & $-$3.4 & 3660-7120 & $R\sim20,000$ & Copernico 1.82m + Echelle \\
2025-11-03 & 60982.89 & $-$3.4 & 3500-7580 & $R\sim400$ & Galileo 1.22m + B\&C + 300tr \\ 
2025-11-04 & 60983.75 & $-$2.5 & 3400-7880 & $R\sim400$ & Galileo 1.22m + B\&C + 300tr \\ 
2025-11-05 & 60984.89 & $-$1.4 & 3430-7880 & $R\sim400$ & Galileo 1.22m + B\&C + 300tr \\ 
2025-11-06 & 60985.79 & $-$0.5 & 3500-8040 & $R\sim450$ & Galileo 1.22m + B\&C + 300tr \\ 
2025-11-07 & 60986.01 & $-$0.3 & 3700-7130 & $R\sim900$ & NOT 2.56m + ALFOSC + gr7 \\
2025-11-07 & 60986.03 & $-$0.3 & 5680-8590 & $R\sim1,300$ & NOT 2.56m + ALFOSC + gr8 \\
2025-11-10 & 60989.01 & $+$2.7 & 3700-7130 & $R\sim900$ & NOT 2.56m + ALFOSC + gr7 \\
2025-11-10 & 60989.02 & $+$2.7 & 5680-8590 & $R\sim1,100$ & NOT 2.56m + ALFOSC + gr8 \\
2025-11-16 & 60996.04 & $+$9.7 & 3400-9720 & $R\sim400$ & NOT 2.56m + ALFOSC + gr4 \\
2025-11-23 & 61002.98 & $+$16.7 & 6190-7840 & $R\sim1,100$ & TNG 3.58m + DOLORES + VHRR \\
2025-11-24 & 61003.01 & $+$16.7 & 3430-8120 & $R\sim350$ & TNG 3.58m + DOLORES + LRB \\
2025-11-25 & 61004.82 & $+$18.5 & 3600-10100 & $R\sim2,100$ & GTC 10.4m + OSIRIS + R2500U/V/R/I \\
2025-12-02 & 61011.05 & $+$24.7 & 3700-9200 & $R\sim300$ & NOT 2.56m + ALFOSC + gr4 \\
2025-12-10 & 61019.89 & $+$33.6 & 3660-7110 & $R\sim20,000$ & Copernico 1.82m + Echelle \\
2025-12-11 & 61020.98 & $+$34.7 & 4870-9100 & $R\sim370$ & Copernico 1.82m + AFOSC + VPH6 \\
2025-12-22 & 61031.02 & $+$44.7 & 3750-8900 & $R\sim400$ & NOT 2.56m + ALFOSC + gr4 \\
2025-12-31 & 61040.93 & $+$54.6 & 3750-8900 & $R\sim300$ & NOT 2.56m + ALFOSC + gr4 \\
2026-01-12 & 61052.74 & $+$66.4 & 4870-8950 & $R\sim250$ & Copernico 1.82m + AFOSC + VPH6 \\
2026-01-23 & 61063.90 & $+$77.6 & 3850-9000 & $R\sim300$ & NOT 2.56m + ALFOSC + gr4 \\
2026-02-16 & 61087.75 & $+$101.4 & 5000-9100 & $R\sim250$ & Copernico 1.82m + AFOSC + VPH6 \\
\hline
\end{tabular}
\tablefoot{The phases reported are relative to the $g$-band first maximum (MJD 60986.3).  ($\star$) Spectra presented by \citet{Taguchi2025ATel17468....1T}. }
\end{table*}

\end{appendix}

\label{LastPage}
\end{document}